\documentclass[a4paper, amsfonts, amssymb, amsmath, reprint, showkeys, nofootinbib, twoside]{revtex4-2}
\usepackage[english]{babel}
\usepackage[utf8]{inputenc}
\usepackage{algorithmic}
\usepackage{graphicx}
\usepackage{siunitx}
\usepackage{booktabs}
\usepackage{physics}
\usepackage{mhchem}
\usepackage{soul}
\usepackage{url}
\usepackage{textcomp}
\usepackage{amsthm}
\usepackage{mathtools}
\usepackage{physics}
\usepackage{xcolor}
\usepackage{graphicx}
\usepackage[left=23mm,right=13mm,top=35mm,columnsep=15pt]{geometry} 
\usepackage{adjustbox}
\usepackage{placeins}
\usepackage[T1]{fontenc}
\usepackage{lipsum}
\usepackage{csquotes}
\usepackage{amsmath}
\usepackage{amssymb}

\usepackage{braket}
\usepackage{natbib}

\def\BibTeX{{\rm B\kern-.05em{\sc i\kern-.025em b}\kern-.08em
    T\kern-.1667em\lower.7ex\hbox{E}\kern-.125emX}}

\DeclareSIUnit\gauss{G}
\DeclareSIUnit\bit{bit}
\DeclareSIUnit \arms {\ensuremath{\mathrm{A_{rms}}}}
\usepackage[pdftex, pdftitle={Article}, pdfauthor={Author}]{hyperref} 
\begin{document}

\title{Optimizing the Electrical Interface for Large-Scale Color-Center Quantum Processors}
\author{Luc Enthoven \textsuperscript{1, 2}}
\email[]{L.A.Enthoven@tudelft.nl}
\author{Masoud Babaie \textsuperscript{1, 3}}
\author{Fabio Sebastiano \textsuperscript{1, 2}}
\email[]{F.Sebastiano@tudelft.nl}
\affiliation{\textsuperscript{1}QuTech, Delft University of Technology, 2600 Delft, GA, The Netherlands}
\affiliation{\textsuperscript{2}Department of Quantum and Computer Engineering, Delft University of Technology, 2600 Delft, GA, The Netherlands}
\affiliation{\textsuperscript{3}Department of Microelectronics, Delft University of Technology, 2600 Delft, GA, The Netherlands}
\date{14-03-2024}




\begin{abstract}
Quantum processors based on color centers in diamond are promising candidates for future large-scale quantum computers thanks to their flexible optical interface, (relatively) high operating temperature, and high-fidelity operation. 
Similar to other quantum-computing platforms, the electrical interface required to control and read out such qubits may limit both the performance of the whole system and its scalability.
To address this challenge, this work analyzes the requirements of the electrical interface and investigates how to efficiently implement the electronic controller in a scalable architecture comprising a large number of identical unit cells.
Among the different discussed functionalities, a specific focus is devoted to the generation of the static and dynamic magnetic fields driving the electron and nuclear spins, because of their major impact on fidelity and scalability.
Following the derived requirements, different system architectures, such as a qubit frequency-multiplexing scheme, are considered to identify the most power efficient approach, especially in the presence of inhomogeneity of the qubit Larmor frequency across the processor. 
As a result, a non-frequency-multiplexed, 1-\SI{}{\milli\meter^2} unit-cell architecture is proposed as the optimal solution, able to address up to one electron-spin qubit and 9 nuclear-spin qubits within a 3-mW average power consumption, thus establishing the baseline for the scalable electrical interface for future large-scale color-center quantum computers.
\end{abstract}


\maketitle

\section{Introduction} 
\label{sec:introduction}
Quantum computers promise significant speedup in solving specific categories of computational problems, such as quantum simulation \cite{feynman2018simulating}, allowing for faster drug discovery and optimization of chemical processes \cite{cao2018potential, cao2019quantum}.
Quantum algorithms can then be executed by operating on the quantum state of quantum bits (qubits), typically by applying and detecting electrical or optical signals to and from the qubits to manipulate and read out their state.
These signals are typically generated at room temperature using off-the-shelf equipment and routed to the qubits usually located at cryogenic temperatures, an approach that suffices for small numbers of qubits.
However, fault-tolerant quantum computing will need $10^3$ - $10^6$ qubits \cite{fowler2012surface}, requiring a more scalable approach to both signal generation and interconnects to improve reliability and cost \cite{Patra2018}.
By using tailored room-temperature (RT) qubit controllers, control over more than 50 qubits has been demonstrated \cite{Arute2019, alberts2021accelerating, bardin2022beyond, zettles202226}.
Nevertheless, for large quantum processors, the required amount of wiring can still lead to an interconnect bottleneck that cannot be solved by any RT controller.
Fortunately, cryogenic electronic controllers, and particularly cryo-CMOS controllers \cite{yoo2023design, Frank2022, Park2021, VanDijk2019_jssc, kang2022cryo} can alleviate the interconnect bottleneck as fewer wires need to enter the cryostat at the expense of additional power dissipated at the cryogenic stage.
Hence, the signal requirements for high-fidelity operation must be well understood such that hardware can be tailored and optimized for performance and low-power operation \cite{VanDijk2019a}.
To optimize cryo-CMOS controllers even further, co-designing the electronics and the quantum processor to define how the qubits are arranged and connected to the controllers can help, for instance, by sharing control signals and circuits for further power reduction.
Consequently, when addressing any qubit platform, investigating the signal requirements and optimizing the controllers for scalability, e.g., for power dissipation and area footprint, are crucial steps to enable larger quantum computers.

Qubits can be implemented in various platforms, each coming with their own advantages and disadvantages.
For instance, qubits based on ion traps have a good qubit-to-qubit connectivity and high operating temperatures but require large voltages, while semiconductor spin qubits offers potential co-integration with the control electronics but the electrical connectivity and the process uniformity of tight-pitch qubit arrays are open challenges \cite{anders2023cmos}.
Among the various quantum-computing platforms, the largest state-of-the-art quantum processors are based on superconducting qubits \cite{Arute2019, chow2021ibm}, which must operate at milli-Kelvin temperatures. 
As current cryogenic refrigerators can offer a very limited cooling power ($\ll$ 1 mW) at those temperatures, the electronics must be placed at a higher-temperature stage, thus posing a stringent interconnect bottleneck between the qubits and the higher-temperature electronics.
Furthermore, these qubits can be entangled only through superconducting couplers, limiting their interconnect capability and posing scalability challenges when connecting multiple quantum processors \cite{storz2023loophole}.

Compared to superconducting qubits, qubits implemented as color centers in diamond relax the constraints on both the operating temperature and the interconnection between the qubits, while also offering high-fidelity control and readout \cite{Ruf2021,pezzagna2021quantum}, making them a promising candidate for future quantum processors.
Their operating temperatures can be higher than \SI{1}{\kelvin}, where typical cryogenic refrigerators can offer significantly more cooling power than at \SI{}{\milli \kelvin} temperatures, increasing the available power budget for cryogenic electronics and hence facilitating qubit/electronics co-integration.
Additionally, since qubit initialization, readout, and entanglement of color centers happen optically, qubit interactions can extend even beyond a kilometer with optical fibers \cite{Hensen2015, Pompili2021, hermans2022qubit}.
Nevertheless, building a large-scale and compact quantum computer that combines several color centers with the required photonic and electronic infrastructure demands complex 3D-integration schemes and techniques that are still being developed \cite{Ishihara2021}. 
Furthermore, the space required by the photonic components will increase the distance between the qubits and the field-generating coils that drive qubit operations, requiring large amplitudes of the electrical driving signals.
Although several cryo-CMOS controllers have been demonstrated, they have been optimized for semiconductor spin qubits, superconducting qubits and ion traps \cite{anders2023cmos}, the design space for diamond-based qubit controllers has been left largely unexplored. 

To bridge this gap, this work first analyzes the requirements of the electrical interface for a vacancy-center-based quantum computer and derives the specifications for individual blocks of the electronic controller. 
With the goal of pursuing a scalable and power-efficient cryo-CMOS controller for vacancy-center-based quantum processors, this paper then goes on to investigate the design trade-offs in the system architecture and implementation, resulting in a comprehensive analysis and leading to an optimized electrical interface.
The paper is organized as follows. 
First, Section \ref{sec:vcqc} summarizes the qubit operations and the related control signals and briefly describes the high-level architecture of a large-scale diamond-based quantum computer.
Then, Section \ref{sec:mf_specs} derives specifications for high-fidelity operations, after which Section \ref{sec:sys_arch} presents a system level controller architecture that can meet these specifications.
Section \ref{sec:sys_impl} proposes a system implementation and estimates its power dissipation.
Finally, Section \ref{sec:conclusion} presents a conclusion of the analysis presented in this paper.

\section{Color Center Based Quantum Processor}

\label{sec:vcqc}

\subsection{Qubits in Color Centers}
\begin{figure}[t]
    \centering
    \includegraphics[width=0.99\linewidth]{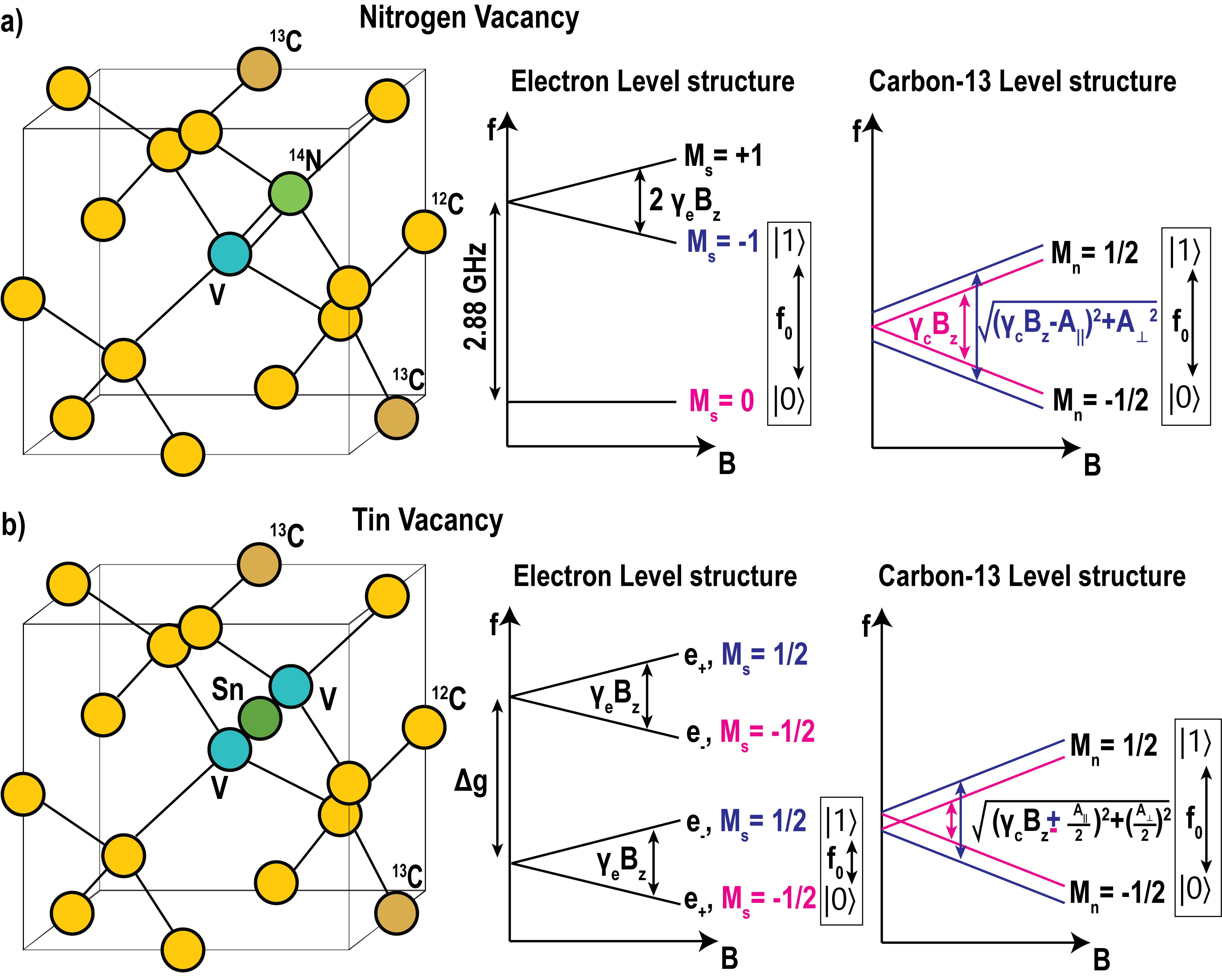}
    \caption{Atomic structure and level structure for the electron and nuclear spins as a function of magnetic field for a) a nitrogen-vacancy center \cite{Bradley2019} and b) a tin-vacancy center \cite{nguyen2019quantum}. The Larmor frequency of the carbon-13 spins depends on the state of the electron spin and their interaction, with the blue and pink levels related to the electron-spin state in blue and pink, respectively. Parameters $A_\parallel$ and $A_\perp$ indicate the coupling of the nuclear spin to the electron spin, $\gamma_e$ and $\gamma_c$ indicate the gyromagnetic ratio of the electron and carbon spin and $B_z$ indicates the magnetic field along the color center axis, these are further discussed in Section \ref{sec:mf_specs}.}
    \label{fig:level_structure}
\end{figure}

Color centers in diamond are formed when a donor atom, e.g., a group-V element like nitrogen or a group-IV element like tin, is implanted or deposited in the diamond and causes a defect \cite{rugar2020generation, luo2022creation}.
The addition of these atoms can cause vacancies in the diamond lattice, leading to nitrogen-vacancy color centers (NV centers) \cite{schirhagl2014nitrogen, manson2006nitrogen} or tin-vacancy color centers (SnV centers) \cite{iwasaki2017tin, rugar2019characterization}.
Throughout this work, SnV centers are used as an example of a group-IV color center, which all have a similar energy-level structure and behave similarly \cite{thiering2018ab, iwasaki2020color}.
Possible atomic structures resulting from the vacancies are shown in Fig. \ref{fig:level_structure}.
The molecular orbital of the color center is then created by the orbitals of the surrounding carbon atoms and the donor atom, giving rise to the different energy levels and properties of the color center.
Additional electrons can be introduced to the molecular orbital by initializing the color center in a charge state, which is typically done by illuminating the color center with a laser \cite{doi2014deterministic, gorlitz2022coherence}.
In literature, color centers are often initialized in their negative charge state, i.e., $\mathrm{NV^-}$, $\mathrm{SnV^-}$, where the color center has absorbed an electron from the environment.
For $\mathrm{NV^-}$, the electrons present will form a spin-1 system (S=1) \cite{Doherty2013}, whereas the electrons in $\mathrm{SnV^-}$, will creates a spin 1/2 system (S=1/2) \cite{Debroux2021}.
In the presence of a magnetic field, the Zeeman effect occurs, causing the energy levels of the spin to be split \cite{acosta2013nitrogen}.
A subset of the available energy levels can then be used to form a qubit for the quantum processor as shown in Fig. \ref{fig:level_structure}, where the qubit states are split by the Larmor frequency ($f_0$).


In addition to the color center, other atoms with magnetic spin are present in the diamond, e.g., \ce{^{14}N} (S=1) or \ce{^{13}C} (S=1/2),
\footnote{\ce{^{13}C} and \ce{^{14}N} appears with typical concentrations of $\approx$1.1\% and $\approx$0.01\% in diamond \cite{mindarava2020synthesis, woods1990nitrogen}.} 
which form a `spin bath' \cite{Bradley2019, Yang2016}.
Since all these spins have interactions and therefore affect each other's Larmor frequency, they will fluctuate over time and cause dephasing, if left uncontrolled.
The variation of the Larmor frequency due to the environment is captured by the $T_2^*$ of the color center, which can be measured with a Ramsey experiment \cite{kobayashi2020electrical}.
Periodic qubit rotations, such as Carr-Purcell-Meiboom-Gill (CPMG) sequences, can help decouple the qubit from its environment and reduce the influence of $T_2^*$ \cite{wang2012comparison, Yang2016}.
While the nearby spins can be a source of decoherence, they can also be used to create additional qubits around the color center since their locations are all static in the diamond lattice \cite{childress2006coherent}.
For instance, a qubit can be defined by the nuclear spin of the \ce{^{13}C} atoms that are naturally present in diamond. 
In that case, the Larmor frequency between the $\ket{0}$ and $\ket{1}$ state of the nuclear spin is defined by both the Zeeman energy and interaction with the electron spin, as shown in Fig. \ref{fig:level_structure}.

While NV center qubits can operate at room temperature, they are often cooled to low temperatures to reveal the fine-level structure of their excited state, which is required to optically entangle distant NV centers \cite{Doherty2013}.
In turn, SnV centers need to operate at cryogenic temperatures to ensure that the electronic spin state is maintained.
This originates from the level structure of SnV centers, which has two spin levels ($M_s=\pm\frac{1}{2}$) and two orbital levels ($e_\pm$), as illustrated in Fig. \ref{fig:level_structure} \cite{Ruf2021}.
When the operating temperature is too high, phonons in the diamond will cause transitions between the orbital states.
As a result, electron spin states are mixed and the electron spin coherence can be lost \cite{Trusheim2020}.
While other group-IV color centers, such as SiV and GeV, require temperatures <\SI{100}{\milli\kelvin}, it is expected that the SnV will have long spin coherence time at temperatures as high as \SI{1}{\kelvin} thanks to the larger splitting of the orbital levels \cite{Ruf2021}.

\subsection{Single-Qubit and Conditional Single-Qubit Rotations}
\begin{figure*}[t]
    \centering
    \includegraphics[width=2 \columnwidth]{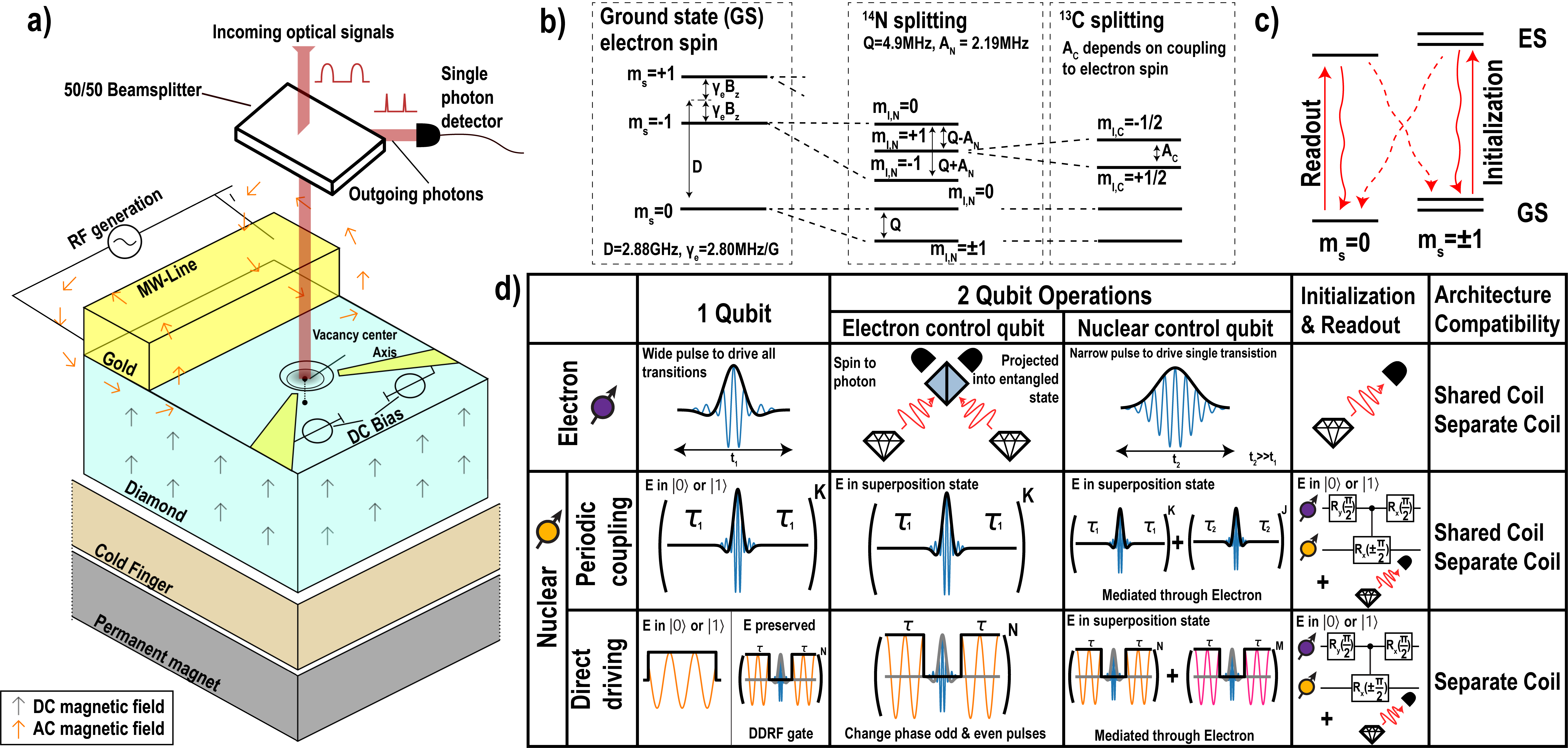}
    \caption{a) Typical free-space setup for nitrogen-vacancy centers in diamond. The diamond is mounted on a cold finger inside a cryostat, while the permanent magnet induces the Zeeman splitting. The MW line introduces an AC magnetic field and the DC bias electrodes tune the electric field to change the wavelength of the emitted photons. A solid immersion lens fabricated around the color center improves the light collection. Optical signal sources and filters are omitted for clarity. b) The NV center hyperfine interactions split the electron spin state energy levels. By driving all energy levels equally, unconditional rotations are performed. Driving a specific energy level results in conditional qubit gates. c) Initialization and readout of the electron spin state occurs through optical transitions between the ground state (GS) and excited state (ES). \cite{thesis_connor}.
    d) Overview of the qubit gates that can be performed for the electron (E) and nuclear (N) spin qubits in NV centers.
    Nuclear spin operations can be performed by either 'periodically coupling' the electron spin to the nuclear spin or by 'directly driving' the Larmor frequency of the nuclear spin.
    Colors represent different frequencies for electron spin (blue) and nuclear spins (orange, pink). $\tau_1$ and $\tau_2$ resonantly couple with different nuclear spins, while $\tau$ decouples the electron from the environment. 
    Two-qubit Operations are conditional rotations where the control qubit is listed in the respective header of the table column, except for the E to E operation where the two qubits are projected in an entangled state through measuring the photons.
    Nuclear spins can be measured and initialized with measurement-based initialization (MBI) or SWAP sequences \cite{thesis_connor, thesis_pfaff}. 
    Architecture compatibility refers to the use of a shared driver and coil or separate driver and coil, which is discussed in Section \ref{sec:sys_arch}. \label{fig:nv_system}}
\end{figure*}

Both unconditional single-qubit rotations and conditional, spin-selective, single-qubit rotations on the different spins in the system can be performed by applying an AC magnetic field at the Larmor frequency, as shown in Fig. \ref{fig:nv_system}(d) \cite{vanderSypen2005}.
The resulting speed of the operation is proportional to the magnitude of the AC magnetic field, and thus fast operations with high Rabi frequencies require large magnetic-field amplitudes.
Typically, Rabi frequencies in the order of \SI{10}{\mega\hertz} are desired, as this results in shorter operations and hence better fidelity for a given qubit coherence time \cite{10_qubit_array_sup,VanDijk2019a}.
Another argument for large Rabi frequencies is the need to perform single-qubit operations on the electron independent on the state of the spin bath.
For instance, the \ce{^{14}N} and \ce{^{13}C} spins have hyperfine interactions with the electron spin in NV centers, changing the Larmor frequency by \SI{\approx \pm 2.5}{\mega\hertz} depending on the spin states, as shown in Fig. \ref{fig:nv_system}(b), such that multiple resonances appear in the spectrum.
Thus, the pulse must drive the different Larmor frequencies equally, requiring a fast pulse, which can then be spectrally flat (e.g., Hermite envelope) with a bandwidth larger than \SI{\approx 10}{\mega \hertz} [see Fig. \ref{fig:nv_system} (d) Electron, 1 Qubit] \cite{Abobeih2018}.
Alternatively, spin-selective (i.e., conditional) qubit rotations can be performed by driving individual Larmor frequencies at low Rabi frequencies and long gate durations as shown in Fig. \ref{fig:nv_system}(d) (Electron, Nuclear control qubit).
Here, pulse shaping can be applied to reduce spectral leakage that drives other spin states.
Most experiments demonstrating coherent control with AC magnetic fields have been performed on NV Centers and can achieve high fidelity \cite{pezzagna2021quantum}. 
Similar qubit operations can be performed on the electron spin of Group-IV color centers by using AC magnetic fields; however, the effective Rabi frequency is expected to be reduced compared to NV centers due to the strain required to mix the orbital states ($e_+$, $e_-$, Fig. \ref{fig:level_structure}) \cite{rosenthal2023microwave}. 

Single-qubit operations on the nuclear spin qubits can be performed by either applying an AC magnetic field resonant to the Larmor frequency [Fig. \ref{fig:nv_system}(d), direct driving] \cite{vanderSypen2005}, which can be used to perform operations independent of the state of the electron spin, or by resonantly coupling the electron spin to the carbon spin [Fig. \ref{fig:nv_system}(d), periodic coupling] \cite{taminiau2014universal}, which disadvantageously requires the electron spin to be in a certain state depending on the operation that needs to be executed.
Depending on the state of the electron spin and the pulse sequence, either conditional gates or unconditional gates can be performed on the nuclear spin [Fig. \ref{fig:nv_system}(d), Nuclear -- Electron Control qubit].
Entanglement between multiple nuclear spins can then be obtained by using the electron spin as mediator and more complicated sequences [depicted in Fig. \ref{fig:nv_system}(d), Nuclear -- Nuclear Control qubit] \cite{taminiau2014universal, Bradley2019}.
The nuclear spin has a lower gyromagnetic ratio, which affects three properties of the nuclear spin qubits:
First, the Larmor frequency required for operations on nuclear spins is much lower than the frequency required for electron spin operations, i.e., \SI{}{\mega\hertz} instead of \SI{}{\giga\hertz}; 
Second, the Rabi frequency is much lower for the same AC magnetic-field amplitude; 
Finally, the nuclear spins are much less sensitive to the magnetic environment compared to the electron, allowing the nuclear spin qubit to serve as a memory qubit with a much longer coherence \cite{Ruf2021}.
In practice, however, the lower Rabi frequency of nuclear spins requires the operations to be so long that the electron spin can decohere during the operation.
To prevent decoherence, more elaborate control sequences, such as dynamically decoupled RF gates [DDRF gate, Fig. \ref{fig:nv_system}(d)] can be employed, where the electron spin is periodically decoupled from the environment to preserve its spin state, while also manipulating the nuclear spin state (in)dependently of the electron spin state by varying the phase of the generated pulses.

\subsection{Qubit Initialization, Readout and Remote Entanglement}

Color centers typically have multiple optical transitions between their ground state (GS) and excited state (ES), which are split by specific energies.
By exciting the color center with a photon that has the energy of a specific transition, the transition can be resonantly excited, causing the electron spin to move to the excited state and emitting a photon when it moves back to the ground state, represented by the solid lines in Fig. \ref{fig:nv_system}(c) \cite{Doherty2013, pezzagna2021quantum, Hensen2015}.
Hence, by exciting specific transitions, the state of the electron spin can be probed due to the presence or absence of a photon.
When calibrated, the color center can decay back towards the same spin state with high probability (i.e., having good cyclicity) \cite{hopper2018spin}.
However, there is also a probability that the electron decays back into a different spin state, as represented by the dashed lines of Fig. \ref{fig:nv_system}(c).
The probability of returning to the same spin state in the ground state depends on the overlap of the eigenstates between the GS and ES, which are influenced by magnetic fields and strain.
Because of the optical losses, multiple excitations of the color center are typically required to ensure an outgoing photon being measured.
Good cyclicity of the transition from the excited state to the ground state is then desired, as a decay into the other spin state reduces readout fidelity \cite{hopper2018spin}.
For initialization of the electron spin qubit, either the spin state can be directly measured, which is susceptible to photon losses similar to the readout, or a different optical transition with low cyclicity can be used such that the electron spin state gets trapped in the other spin state [dashed lines, Fig. \ref{fig:nv_system}(c)], enabling higher fidelity initialization at the expense of using another transition and hence an additional laser wavelength \cite{thesis_connor}.
Initialization and readout of the individual nuclear spin states are achieved by entangling the nuclear spin with the electron spin, whereas pulse polarization sequences have shown polarization and hence initialization of the nuclear spin bath through the electron spin \cite{schwartz2018robust}.

Remote entanglement between the electron spin of different color centers can be created by using their emitted photons, which are entangled with the spin state.
To create the entangled state, two photons from different color centers need to have identical wavelengths and the photons need to pass through a 50/50 beam splitter to make it impossible to identify the specific qubit generating each photon, after which the outputs are measured, as shown in Fig \ref{fig:nv_system}(d) Electron -- Electron control qubit \cite{bernien2013heralded}.
Depending on the photon measurement outcomes, the two color centers are projected in an entangled state \cite{thesis_pfaff}. 
Due to the probabilistic nature of the measurement, not all attempts result in effective entanglement.
Furthermore, generating identical photons is probabilistic, as photons coming from the color center are either emitted in the zero-phonon line (ZPL) or the phonon sideband (PSB).
For creating entanglement, only photons coming from the ZPL can be used as they have a narrow emission spectrum.
However, color centers are sensitive to strain and magnetic fields, which affect the wavelength of ZPL photons and requires the wavelengths to be tuned.
In addition, the excited states of NV centers are also sensitive to electric fields, which allows tuning the ZPL by applying DC electric fields as illustrated in Fig. \ref{fig:nv_system}(a) \cite{sipahigil2014indistinguishable}.
One of the main reasons to move towards group-IV centers, such as the SnV, is their ability to emit more photons in the ZPL compared to NV centers, meaning that entanglement between distant color centers can be generated at higher rates \cite{Ruf2021}.

\subsection{Vacancy-center-based quantum processor}
\begin{figure*}[t!]
    \centering
    \includegraphics[width=2 \columnwidth]{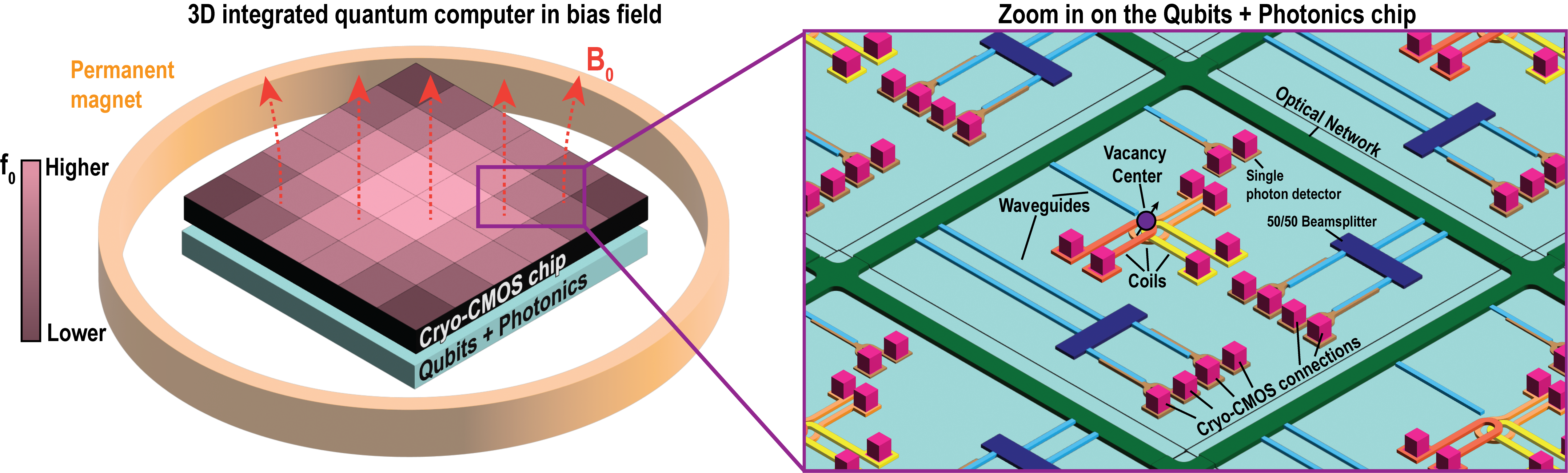}
    \caption{Illustration of a quantum processor based on color centers in diamond, showing the 3D-integrated cryo-CMOS chip, the qubits and the photonics on the left and the components present on the photonics chip on the right.}
    \label{fig:nv_scalable}
\end{figure*}

Fig. \ref{fig:nv_system}(a) depicts the typical free-space setup used for NV center experiments and summarizes the functionality required by the control interface.
The color center is cooled to a temperature below a few Kelvin and a DC magnetic field creates the Zeeman splitting that defines the Larmor frequency of the qubits. 
For the electronic control, magnetic fields with different frequencies are required to perform operations on the electron and nuclear spin.
These fields are generated by large currents running in metallic striplines to obtain high Rabi frequencies. 
For the optical control, at least two lasers with different wavelengths are needed to initialize the color center in the correct charge state and to initialize and read out the color center electron spin state.
For the readout and entanglement, single photons need to be detected by single-photon detectors.

Although a complex setup may be required, all those functionalities can be routinely implemented in an experimental lab environment.
However, these setups only host a few qubits, which is limited by the interconnect bottleneck and the complexity of the optical and electronic controllers, as mentioned in Section \ref{sec:introduction}.
To move towards the goal of large-scale quantum computers, an integrated approach to address the scalability limitations has been proposed.
Here, multiple color centers can be combined on a single photonic chip that is 3D-integrated with a CMOS IC, comprising of circuits for qubit biasing, qubit control, and for controlling the photonic components \cite{Ishihara2021}.
The quantum processor will consist of identical unit cells, each hosting a color center together with the required optical/electrical signals for qubit control. 
While each color center has many spins in its environment that can be used to create qubits, in the proposed unit cell each color center will host 10 qubits, i.e., 1 electron spin qubit and 9 nuclear spin qubits, such that there is sufficient addressability of the adopted qubits that are present.
Ideally, the computing power of such tiled quantum processor could be increased by simply adding more unit cells, with the number of unit cells only limited by the allowed size and the cooling power available from the cryostat. 
Hence, minimizing the area and the power dissipation of the unit cell has a direct impact on the scalability of this approach.
Fig. \ref{fig:nv_scalable} illustrates such vision with the unit cell on the photonic chip hosting the coils for generating the driving magnetic fields, the photon detectors, the waveguides, the beams splitters and the optical network, with most components wired to the driving CMOS circuits via 3D interconnects.
The coils drawn in the figure enable inducing magnetic fields in different orientations and have non-negligible distance to the qubit due to the presence of photonic components that interface with the color center, thus requiring large currents for a certain magnetic field.
In the drawn scenario, each unit cell has three individual magnetic-field-inducing coils that are connected to a dedicated CMOS controller and driver, which is further elaborated upon in Section \ref{sec:sys_arch}.
Furthermore, the whole quantum processor is biased with a permanent magnet to set the Larmor frequency of the unit cells.
If the unit cells become too large, the position of the color centers will span a wide area, hence being more subject to any inhomogeneity in the bias magnetic field.
Thus, the unit-cell size must be kept just large enough to fit all the electronics and optical components, i.e., indicatively in the order of \SI{1}{\milli\meter} $\times$ \SI{1}{\milli\meter}, but not larger than that.

In the following, we derive the specifications of the controller by focusing on the unit-cell architecture shown in Fig. \ref{fig:nv_scalable}. Note, however, that the specification study reported below can be generally applied to any color-center qubit and to other present and future quantum-processor architectures.

\section{Specifications of the electronic interface}
\label{sec:mf_specs}

Previous work on deriving the specifications for the controller of spin qubits in semiconductors is used as a starting point for deriving the specifications of the AC magnetic field \cite{VanDijk2019a, VanDijk2020}, since the expressions for the fidelity remain the same under the assumption that the color center is a 2-level system.
This assumption holds if the nuclear spins are properly initialized or if the MW pulses address the different nuclear-spin states equally.
While all the general analysis is presented, we also report a specific numerical example to convey to the reader an order of magnitude of the requirements. Here, a fidelity of 99.99\% for both operations through the AC field and idling is targeted, with the error budget divided (arbitrarily) equally among 8 components for the operations and among 4 components for the idling, which result in the requirements of Table \ref{tab:specifications}.

\subsection{DC Magnetic Field}
For the DC magnetic environment, one can distinguish between the magnetic field parallel ($B_\parallel$) and orthogonal ($B_\perp$) to the axis along which the vacancy and the substitutional atom are located (Fig. \ref{fig:level_structure}).
Typically, biasing fields are aligned to the NV center or SnV center axis as other terms in the Hamiltonian, such as the zero-field splitting, can be also along this axis \cite{Doherty2013}.

\subsubsection{Parallel Magnetic field Requirements $B_\parallel$}
The parallel DC magnetic field $B_\parallel$ determines the Larmor frequency for the various qubits around the color center.
While the electronic level structures of the electron spin in NV and SnV centers can be complex \cite{Doherty2013, rosenthal2023microwave}, the Larmor frequency of the NV centers (SnV centers) due to a magnetic field along $B_{\parallel}$ can be estimated as:
\begin{equation} \label{eq:Larmor}
f_{0,e,NV}=|D-\gamma_{e} B_{\parallel} | \hspace{2em} (f_{0,e,SnV}=|\gamma_{e} B_{\parallel}|).
\end{equation}
where $D$=\SI{2.88}{\giga\hertz} is the NV's zero-field splitting defined along the Nitrogen-vacancy axis and $\gamma_{e}$=\SI{2.8}{\mega\hertz\per\gauss} is the electron gyromagnetic ratio (Fig. \ref{fig:level_structure}).
For nuclear spin qubits, the Larmor frequency largely depends on the magnetic field, but also depends on the state of the electron spin ($m_s$), as shown in Fig. \ref{fig:level_structure}.
In case of NV centers, the Larmor frequency for the carbon-13 spins is given by $f_{0,c} = |\gamma_c B_z|$ for $m_s = 0$ and $f_{0,c} = \sqrt{(\gamma_c B_z-A_\parallel)^2+A_\perp ^2}$ for $m_s = -1$, with $\gamma_c$=\SI{1.0}{\kilo\hertz\per\gauss} the carbon gyromagnetic ratio, and $A_\parallel$ and $A_\perp$ the parallel and perpendicular hyperfine interactions of the nuclear spin with the electron spin \cite{Bradley2019}.
For SnV centers, the carbon-13 spin Hamiltonian suggests that $f_{0,c} = \sqrt{(\gamma_c B_z \pm (A_\parallel/2))^2+(A_\perp/2) ^2}$, where the sign depends on the electron being in $m_s = +\frac{1}{2}$ or $m_s = -\frac{1}{2}$ \cite{metsch2019initialization}.
The exact values of $A_\parallel$ and $A_\perp$ depend on the location of the carbon nuclear spin with respect to the color center, and values between \SI{10}{\kilo \hertz} and \SI{100}{\kilo \hertz} are reported in the literature \cite{10_qubit_array_sup}.

For all target qubits, the Larmor frequency should be significantly higher than the Rabi frequency to allow for high-fidelity operations \cite{VanDijk2019a}.
Furthermore, using strong magnetic fields (\SI{\geq1800}{\gauss}) is preferred to reduce the decoherence of carbon spins during remote entanglement generation \cite{kalb2018dephasing, Pompili2021}.
However, larger magnetic fields will also require higher frequencies of the AC magnetic field, which also need to be generated by the controller.
Hence, DC magnetic fields between \SIrange[]{2000}{10000}{\gauss} are expected to enable both high fidelity and practical frequencies for the controller.
Even though the Larmor frequency of the color center can be set and calibrated, some residual inaccuracy in the Larmor frequency can appear due to a finite frequency resolution in the AC driver or a finite resolution in the magnetic-field control.
As a result, a slowly accumulating error in the tracked frequency causes infidelity when idling:
\begin{equation}\label{eq:inf_df}
    1-F=1-\cos^2\left(\frac{\Delta\omega T_{op}}{2}\right),
\end{equation}
where $F$ is the fidelity, $\Delta\omega$ is the inaccuracy of the tracked frequency in rad/s and $T_{op}$ is the operation time during which the qubit idles \cite{VanDijk2019a}.
This frequency error can be reduced by tuning the DC magnetic field, or by changing the tracking frequency of the qubit to limit $\Delta\omega$.
In Table \ref{tab:specifications}, the correction by tuning the DC magnetic field is assumed, resulting in a required magnetic-field resolution of \SI{5.7}{\milli\gauss} for a \num{2.5e-5} infidelity, corresponding to a frequency error of \SI{15.9}{\kilo\hertz} for the electron spin qubit when idling.

In addition to the static inaccuracy, slow fluctuations in the spin bath and magnetic noise around the color center will introduce fluctuations in the magnetic field.
Consequently, the Larmor frequency will vary and lead to a limited fidelity, as it causes an error in phase tracking.
Approximating the slow magnetic field noise with a static error, the infidelity is \cite{VanDijk2019a}:
\begin{equation}\label{eq:inf_t2star}
    1-F=\frac{1}{4}(\omega_2^*)^2T_{op}^2.
\end{equation}
where $\omega_2^* = \frac{\sqrt{2}}{T_2^*}$.

For faster noise fluctuations, one needs to consider the power spectral density of the magnetic field noise.
The fluctuating spins from the spin bath can be modeled in more detail by assuming them to be an Ornstein-Uhlenbeck process \cite{Yang2016}.
The value of $T_2^*$ together with information on the noise auto-correlation time $\tau_c$ can be used to describe the power spectral density (PSD) of the Larmor frequency of the qubit \cite{de2010universal}:
\begin{equation}\label{eq:noise_cor}
    S(\omega ) = 2\pi (\omega_2^*)^2\frac{\frac{1}{\tau_c \pi}}{\omega ^2 + (\frac{1}{\tau_c})^2}.
\end{equation}
The operations performed on the qubit serve as a noise filter function $H_n(\omega)$.
Without any operations applied to the color center, the noise filter function is given by
\begin{equation} \label{eq:noise_bw}
    |H(\omega)|^2 = \frac{\sin(\frac{T_{op}\omega}{2})^2}{\omega^2},
\end{equation}
With this, the infidelity caused by the noise can be estimated with \cite{Green2013}:
\begin{equation}\label{eq:noise_inf}
    1 - F_{noise} = \frac{1}{\pi}\int_{0}^{\infty}S^2(\omega)|H_n(\omega)|^2 d\omega.
\end{equation}
Consequently, depending on type of noise, a simplified [Eq. (\ref{eq:inf_t2star})] or more complicated noise model [Eq. (\ref{eq:noise_inf})] can be used.
When the noise present is slowly fluctuating, which typically is the case for a spin bath, one can consider the noise to not change during qubit operation, such that (\ref{eq:inf_t2star}) provides sufficient accuracy and yields the same result as (\ref{eq:noise_inf}).
Nevertheless, due to the high frequency noise and interference from the electronics, eq. (\ref{eq:noise_inf}) provides more accurate results.
For simplicity, the noise reported in Table \ref{tab:specifications} is assumed to be white and by using the ENBW of eq. (\ref{eq:noise_bw}), the budgeted infidelity can be converted to a PSD with equation eq. (\ref{eq:noise_inf}).

Any operation performed on the qubit will affect its noise filter function.
By periodically applying operations to dynamically decouple the qubit from the spin bath, it is possible to reduce the infidelity the spin bath causes \cite{wang2012comparison}.
The dynamical decoupling sequences change the noise filter function previously discussed, as periodic qubit rotations around the X/Y axis cancel out the accumulated phase \cite{Yang2016}.
While this reduces the sensitivity of the qubit to low-frequency noise in the environment, it does require additional operations and increases the sensitivity to noise around the dynamical decoupling frequency.
Since this offers several options to minimize the effect of noise and it may be algorithm dependent, the qubit is assumed to be idling for the estimations reported here.

\subsubsection{Orthogonal Magnetic field requirements $B_\perp$}
Similar to the magnetic field in parallel to the color center, the noise on the orthogonal magnetic field can also cause infidelity.
However, the filter function changes compared to the parallel magnetic field and is described as \cite{VanDijk2019a}:
\begin{equation} \label{eq:noise_bw_xy}
    |H(\omega)|^2 = 2\frac{\sin(\frac{T_{op}}{2}(\omega-\omega_0))^2}{(\omega-\omega_0)^2},
\end{equation}
where $\omega_0 = 2\pi f_0$.
The noise on the orthogonal magnetic fields must then be limited around the Larmor frequency.
The high gyromagnetic ratio of the electron spins causes the filter function to be centered around GHz frequencies, which can be easily filtered. 
Nuclear spins have a filter function centered in the range of \SIrange{2}{10}{\mega \hertz} due to their lower gyromagnetic ratio, which are more difficult to filter.
To compute the specification for Table \ref{tab:specifications}, we use (\ref{eq:noise_inf}) with the infidelity budget, resulting in a requirement for the PSD of $B_\perp$ around $\omega_0$ for both electron- and carbon spin qubits when idling.

\begin{figure}[t]
    \centering
    \includegraphics[width=0.95\linewidth]{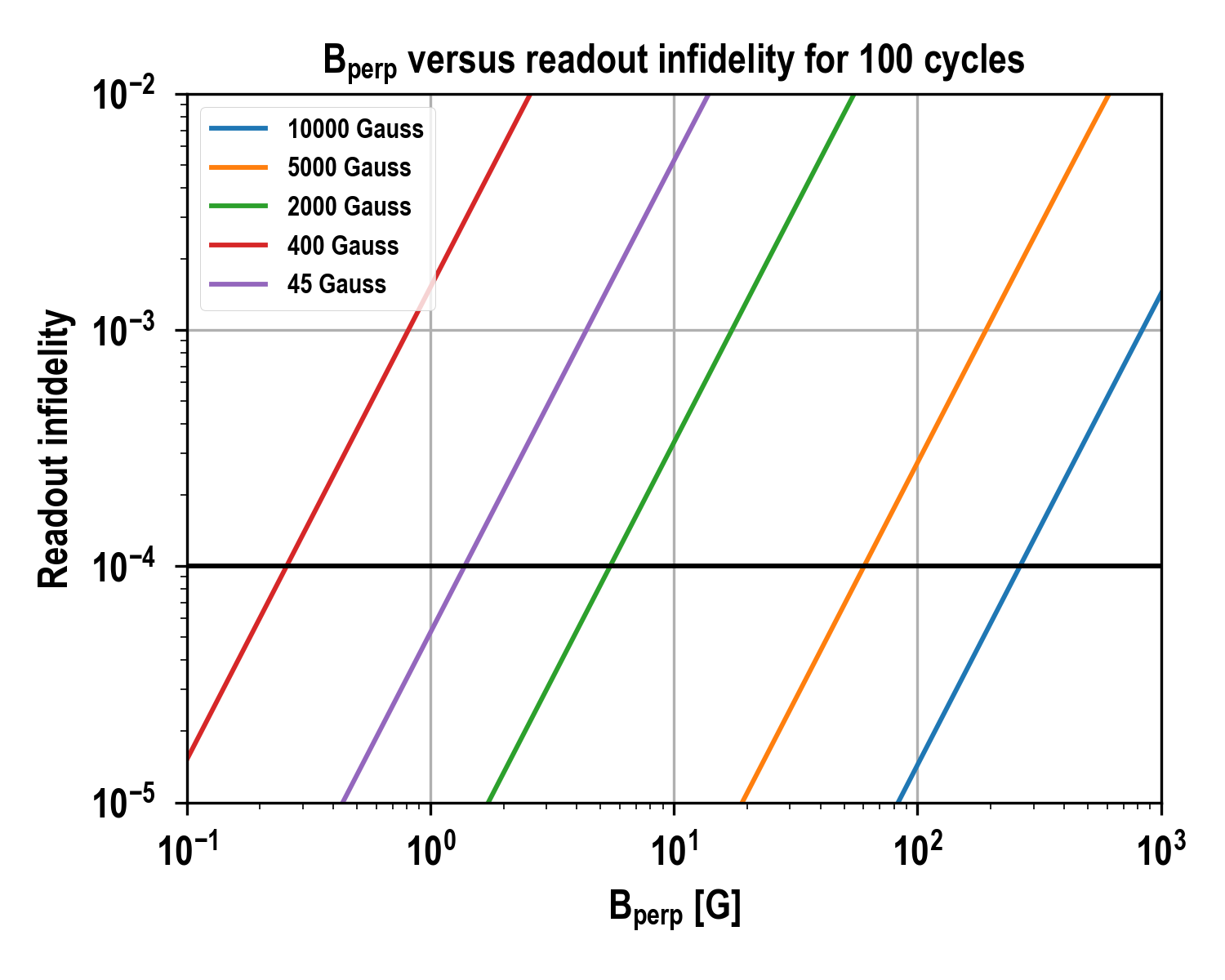}
    \caption{Readout infidelity of an NV center versus perpendicular magnetic fields ($B_\perp$) for various permanent magnet strengths, computed by taking the overlap between the excited and ground state Hamiltonian's, assuming no strain and $N=100$ readout cycles. }
    \label{fig:readout_vbperp}
\end{figure}

The perpendicular DC field also influences the eigenstates of the spins in the system, changing the cyclicity of the readout transition and affecting the readout fidelity.
During readout, the electron spin moves from the ground state to the excited state, where it is subject to the influences of strain as discussed in Section \ref{sec:vcqc}, while there is also a different zero-field splitting present \cite{Doherty2013}.
The presence of additional orthogonal magnetic field $B_\perp$, for example due to misalignment of the magnet with the color center axis, needs to be limited as this reduces the overlap of the eigenstates, lowering the cyclicity of the transition and degrading the readout fidelity.
The readout infidelity due to spin mixing can be computed by $1-F_r = 1-p_{ov}^N$, where $N$ is the number of readout attempts required to collect a photon and $p_{ov}$ is the overlap between the excited state and ground state, giving the probability to decay back into the same spin state.
By assuming a parallel magnetic field, a given strain in the diamond and a required number of readout cycles, requirements for the allowed perpendicular magnetic field can be determined.
Fig. \ref{fig:readout_vbperp} shows the readout infidelity due to the perpendicular magnetic field for various parallel magnetic fields for an NV center.
This is computed by taking the overlap between the low-temperature excited state and ground state Hamiltonian as found in \cite{Doherty2013}, assuming no strain and $N=100$.
For a given $B_\perp$, the readout infidelity first increases when increasing the magnetic field, i.e., from \SI{45}{\gauss} to \SI{400}{\gauss}, and then decreases when moving to even higher fields \SI{>2000}{\gauss}.
This relates to the Larmor frequency of the NV center, which first decreases and then increases again, according to (\ref{eq:Larmor}).
With these assumptions and when allowing a contribution of \num{1e-4} to the readout infidelity due to spin mixing, the perpendicular magnetic field needs to be limited to \SI{5.5}{\gauss} for a parallel magnetic field of \SI{2000}{\gauss} and 100 readout cycles.


\subsection{AC Magnetic Field}
To perform qubit rotations on the electron and nuclear spins, an AC current is required to generate an AC magnetic field perpendicular to the vacancy-center axis with the Larmor frequency of the target qubit.
With magnetic fields in the range of \SIrange{2000}{10000}{\gauss}, the Larmor frequency of the electron spin is between \SIrange{2}{28}{\giga \hertz} and the nuclear spin between \SIrange{2}{10}{\mega \hertz}.

The speed of the qubit operations depends on the gyromagnetic ratio and the strength of the AC magnetic field.
For NV center (SnV center), the Rabi frequency can be computed through 
\begin{equation}\label{eq:Rabi}
f_{r,e} = \left|\frac{1}{\sqrt{2}}\gamma_e B_{ac}\right| \hspace{2em} (f_{r,e} = \left|\eta \gamma_e B_{ac}\right|)
\end{equation}
where $B_{ac}$ is the amplitude of the magnetic field and $\eta$ is the reduction in Rabi frequency due to the orbital mixing for SnV centers \cite{rosenthal2023microwave}.
For the \ce{^{13}C} spins in the environment, the Rabi frequency is $f_{r,c} = |\gamma_c B_{ac}|$ when directly driving \cite{thesis_connor}, while for gates that use periodic coupling the operations speed depends on the interaction strength with the electron \cite{taminiau2014universal}.
In order to achieve $\pi$-rotations on the electron spin from \SIrange{1}{0.1}{\micro \second}, one needs a $|B_{ac}|$ from \SIrange{0.5}{5}{\gauss}. 
Carbon qubit operations happen on longer timescales due to the lower gyromagnetic ratio, thus requiring $|B_{ac}|$ from \SIrange{0.9}{9}{\gauss} for directly driving a $\pi$ rotations within \SIrange{1}{0.1}{\milli \second}.

Specifications for the frequency inaccuracy, frequency noise, phase inaccuracy, timing inaccuracy, timing jitter, amplitude inaccuracy, amplitude noise, and wideband additive noise of the AC signals are computed with the equations presented in Table 1 of \cite{VanDijk2019a} for both nuclear and electron qubits.
Here, all noises except the wideband noise are assumed to apply on longer timescales than the operation time, such that they can be considered as a random static error during the operation, similar to assumptions made in Table 2 of \cite{VanDijk2019a}.
The reported values for the electron spin qubit of the NV center are similar to the previously reported values of \cite{VanDijk2020}, which targets semiconductor spin qubits and superconducting qubits, while the differences between electron- and nuclear-spin qubit requirements originate from the difference in gyromagnetic ratio.

Finally, spurs could drive a rotation on an untargeted qubit, resulting in infidelity, thus requirements should be placed to limit the spurious-free dynamic range (SFDR). 
If the spur is present at exactly the Larmor frequency, the infidelity can be computed with \cite{VanDijk2019a}:
\begin{equation}\label{eq:spur_driving}
    1-F=\frac{1}{4}\omega_{spur}^2T_{op}^2, 
\end{equation}
where $\omega_{spur}$ is the amplitude of the spur that can be converted to a magnetic field.
Alternatively, if a tone is not exactly at the Larmor frequency of the qubit, but slightly detuned, the infidelity can be computed using \cite{VanDijk2019a}:
\begin{equation}\label{eq:fdma_driving}
    1-F\approx\frac{\beta^2}{\alpha^2}\sin^2\left(\frac{\theta}{2}\alpha\right), 
\end{equation}
where $\alpha = \frac{\omega_{0,space}}{\omega_{R,addr}}$, $\beta = \frac{ \omega_{R,unaddr}}{\omega_{R,addr}}$, $\omega_{0,space}$ is the frequency spacing of the tone with the Larmor frequency, $\omega_{R,addr}$ is the Rabi frequency of the addressed qubit, $\omega_{R,unaddr}$ is the Rabi frequency of the unaddressed qubit and $\theta$ is the rotation angle of the targeted qubit \cite{VanDijk2019a}. 
In the unit cell, the electron spin qubit has no other qubit Larmor frequencies close by as the carbon spins are far detuned.
Nevertheless, spurs on the Larmor frequency of the electron spin qubit should be avoided since they will contribute to infidelity when idling.
Based on the targeted operation times of Table \ref{tab:specifications}, spurious tones must be limited to below \SI{8}{\milli \gauss_{pk}}.
Similarly, for the nuclear spins, spurs on the Larmor frequency should be limited to \SI{14.9}{\milli \gauss_{pk}}.
However, an additional consideration applies for nuclear spins, since their Larmor frequencies are very similar and only differentiated by their interaction with the electron spin.
Hence, when driving an operation on a nuclear spin that has a similar Larmor frequency, a rotation can be caused on the untargeted nuclear spin, causing infidelity according to (\ref{eq:fdma_driving}). 
Since all nuclear spins are driven by the same coil, $\beta = 1$, and the introduced infidelity is only dependent on $w_{0,space}$, which in turn depends on the location of the \ce{^{13}C} spins with respect to the color center.
Reducing the crosstalk then implies selecting a set of nuclear spins that are suitably spaced, i.e., $\omega_{0,space} > \omega_{R}$, together with adequate pulse shaping \cite{VanDijk2019a}.
If further reduction of crosstalk is required, the Rabi frequency and hence the amplitude of the driving signal must be reduced.

An example set of specifications for a NV center electron spin and carbon nuclear spin to achieve a 99.99\% fidelity is reported in Table \ref{tab:specifications}.
For this table, a \SI{2000}{\gauss} DC magnetic field sets the Larmor frequency, which is a typical setting for contemporary NV center setups \cite{Bradley2019,hermans2022qubit}.
Furthermore, Rabi frequencies of \SI{5}{\mega \hertz} and \SI{5}{\kilo \hertz} are targeted for the electron and nuclear spin, respectively.
This allows for sufficiently fast operation time $T_{op}$ for $\pi$-rotations, while not requiring excessively large magnetic fields.
The specifications for control electronics to drive high-fidelity qubit gates on the electron spin are similar to the ones previously described in \cite{VanDijk2019a}.
The main difference is for the carbon spins, which, due to their lower gyromagnetic ratio, operate on much longer timescales, are less sensitive to magnetic field noise, and require finer frequency resolution to achieve high fidelity operation.

\begin{table}[t]
    \centering
    \caption{Specifications to achieve 99.99\% fidelity on an NV center for \SI{2000}{\gauss} Field. Target Rabi frequency of \SI{5}{\mega \hertz} for electron operations and \SI{5}{\kilo \hertz} for nuclear operations. For calculating the fidelity, a $\pi$ rotation using a rectangular envelope for both electron and nuclear spin operations is assumed. The individual components each contribute equally to the total infidelity, which is budgeted to 99.99\%.}
    \label{tab:specifications}
    \setlength{\tabcolsep}{3pt}
    \resizebox{\linewidth}{!}{
    \begin{tabular}{l l l r}
        \toprule
        \textit{Target qubit} & Carbon (C) & Electron (E) & Eq. \\ 
        \midrule 
        \textbf{AC Magnetic field} \\
        \midrule 
        Target excitation frequency & \SI{2.1}{\mega \hertz} & \SI{2.7}{\giga \hertz} & (\ref{eq:Larmor}) \\
        Frequency inaccuracy & \SI{17.7}{\hertz} & \SI{17.7}{\kilo \hertz} & (\ref{eq:inf_df}) \\
        Frequency noise & \SI{17.7}{\hertz_{rms}} & \SI{17.7}{\kilo \hertz_{rms}} & (\ref{eq:inf_t2star}) \\
        Phase inaccuracy & \SI{0.20}{^\degree} & \SI{0.20}{^\degree} & \cite{VanDijk2019a}, Tab. I \\
        Duration inaccuracy & \SI{0.23}{\micro \second} & \SI{0.23}{\nano \second} & \cite{VanDijk2019a}, Tab. I \\
        Timing jitter & \SI{0.23}{\micro \second_{rms}} & \SI{0.23}{\nano \second_{rms}} & \cite{VanDijk2019a}, Tab. I  \\
        AC Field amplitude & \SI{4.7}{\gauss} & \SI{2.5}{\gauss} & (\ref{eq:Rabi}) \\
        Amplitude inaccuracy & \SI{11}{\milli \gauss} & \SI{5.7}{\milli \gauss} & \cite{VanDijk2019a}, Tab. I \\
        Amplitude noise & \SI{11}{\milli \gauss_{rms}} & \SI{5.7}{\milli \gauss_{rms}} & \cite{VanDijk2019a}, Tab. I \\
        Wideband additive noise & \SI{8.9}{\milli \gauss_{rms}} & \SI{3.4}{\milli \gauss_{rms}} & \cite{VanDijk2019a}, Tab. I \\
        Max spurious tone @ $f_0$ & \SI{14.9}{\milli \gauss_{pk}} & \SI{8}{\milli \gauss_{pk}} & (\ref{eq:spur_driving}) \\ 
        \midrule 
        \textbf{DC Magnetic field} \\
        \midrule 
        Z-Field accuracy & \SI{15}{\milli \gauss} & \SI{5.7}{\milli \gauss} & (\ref{eq:inf_df}) \\
        Z-Field noise & \SI{44}{\nano \gauss ^2 /\hertz} & \SI{6.4}{\pico \gauss ^2 /\hertz} & (\ref{eq:noise_inf}) \\
        Allowed X/Y Field & - & \SI{5.5}{\gauss} & Fig. \ref{fig:readout_vbperp} \\
        Combined X/Y Field noise & \SI{22}{\nano \gauss ^2 /\hertz} & \SI{3.2}{\pico \gauss ^2 /\hertz} & (\ref{eq:noise_inf}) \\
        \bottomrule
    \end{tabular}
    }
\end{table}

\subsection{Electronic interface for the photonic components}
In addition to qubit biasing and qubit control via the generation of magnetic fields, the electronic interface must also drive the integrated photonic circuitry.
The exact specifications depend heavily on the specific implementations of the photonic components.
While the electronics can influence the operation speed or the system functionality, it does not directly affect fidelity.
Nevertheless, some more generic requirements can be listed, especially for the photon detector but also for other photonic components.

\subsubsection{Photon detectors}
The photon detectors is used to measure the arrival of photons to determine the spin state of the vacancy-center qubits, as described in Section \ref{sec:vcqc}.
Consequently, non-idealities of the photon detector can translate into readout infidelity.
For example, if a photon detector triggers in a given time period resulting in the measurement of the $\ket{1}$ state, this may not necessarily originate from a photon emitted by the color center, but can also be due to background light, or a dark-count event.
Similarly, no photon being measured results in measuring $\ket{0}$, but, in addition to correctly corresponding to the color center not emitting a photon, this can be caused by a detection failure due to limited detection efficiency or by spin-state flipping.
All the mentioned error sources, i.e., background light, non-zero dark counts, limited detection efficiency, and state loss, can limit the readout fidelity. While a readout fidelity >99\% can already be achieved in free-space systems \cite{hermans2022qubit}, the fidelity is expected to further improve when moving to integrated photonics thanks to the expected better detection efficiency.

The single-photon detector has arguably the main influence on the readout fidelity together with the optical losses in the system.
Among the possible choices for photon detectors, superconducting nanowire single-photon detectors (SNSPD) are an attractive choice since they offer low dark-count rates, good detection efficiency, and sufficiently low jitter \cite{chang2021detecting}.
Furthermore, the operating temperature of the color centers is compatible with the SNSPD's typical operating range (<2 K \cite{chang2019multimode}), allowing the 3D-integration of the photonics, including the SNSPDs, with their cryo-CMOS control electronics \cite{ravindran2020active}.
Biasing the SNSPD should not require significant power dissipation, since they are typically biased with low currents (\SI{<20}{\micro\ampere}), while sufficient resolution (\SI{\approx100}{\nano\ampere}) should allow biasing the SNSPD at the highest system detection efficiency.
The amplifier reading out the SNSPD output should ensure that the readout fidelity is not degraded, for instance, by missing counts or incorrectly triggering, while also dissipating little power.
Finally, the photon detector and readout electronics should allow for sufficiently high count rates, which may be required in calibration procedures of the quantum processor. 
Unlike widespread SNSPD sensing systems that are optimized for extremely low jitter, this application may only request moderate jitter performance, e.g., to allow time-filtering of the detected photon to discriminate false positives, thus exploiting the SNSPD's low jitter and allowing for a low-power moderate-noise readout electronics. 

\subsubsection{Other support electronics}
Electrical control may be required by other photonic components, such as optical switches, variable optical attenuators (VOA), interferometers, and possibly devices to strain-tune the color center \cite{bogaerts2020programmable,Ishihara2021}.
Among those components, the VOA and strain-tuning will be calibrated during the quantum-processor bring-up sequence and only require a fixed biasing during the algorithm execution.
Conversely, the switches and interferometers should operate at the quantum-processor operating speed, i.e., at a rate in the order of \SI{100}{\nano\second} based on the Rabi frequency of the electron spin qubit in Table \ref{tab:specifications}.
Although those switching speeds could be easily achievable with cryo-CMOS integrated circuits, a more challenging requirement comes from the required voltage levels for both the statically and dynamically driven components, with the optical switches requiring, for instance, up to \SI{20}{\volt} excitations \cite{seok2016large}.

\section{System Architecture}\label{sec:sys_arch}
As will be clear in the following, the generation of the AC and DC magnetic fields according to the specifications in Table \ref{tab:specifications} will require a significant fraction of the power dissipation of the whole controller. 
Thus, an optimal power-efficient implementation of the magnetic-field drivers in each unit cell is necessary to maximize the number of unit cells in the processor.
Prior art for qubit drivers assumes either a dedicated AC/MW line for each qubit with a dedicated qubit driver or a single line serving multiple qubits through a shared driver via frequency-division multiple access (FDMA) \cite{Frank2022, Park2021}. 
Since no scalable cryo-CMOS driver or controller has been yet demonstrated for vacancy-center quantum processors, the advantages and disadvantages of a shared driver or a dedicated driver will be considered and trade-offs will be listed, with the goal of selecting the most scalable approach.

\subsection{Shared driver}
\begin{figure*}[t!]
    \centering
    \includegraphics[width=2 \columnwidth]{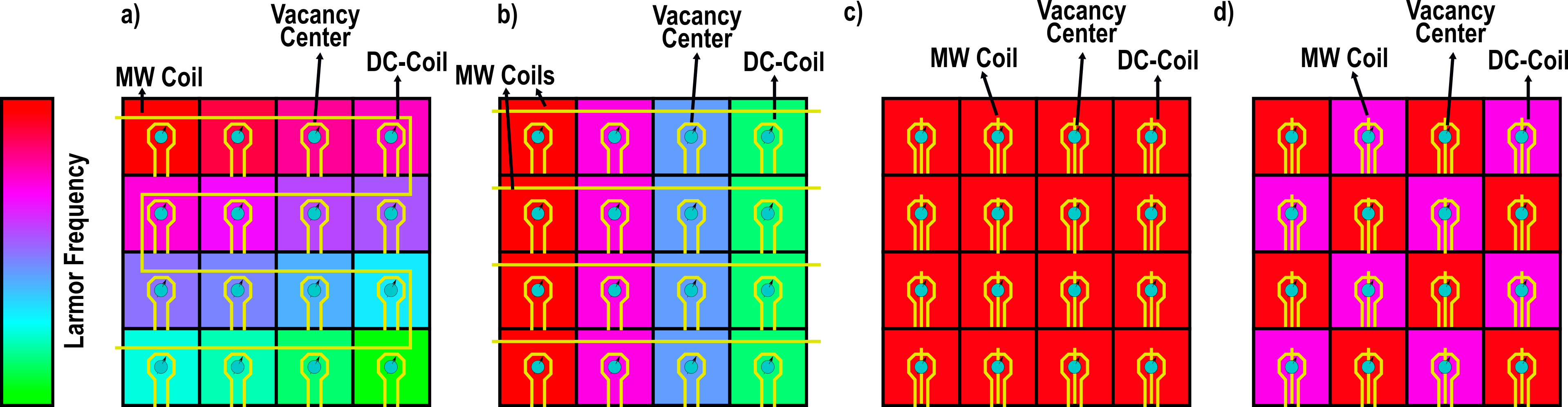}
    \caption{Unit-cell architectures for multiplexing the AC driving signal: a) A shared coil distributes the MW fields to all the color centers, whose Larmor frequencies are spaced by using local magnetic biasing via the DC-coils; b) Similar to a), but with multiple MW coils, each addressing a subset of the unit cells; c) An individual MW coil is used per unit cell, removing the need for frequency spacing; d) Dedicated MW coils per unit cells are employed as in c) but a checkerboard pattern facilitated by local DC-coil biasing is employed for the Larmor frequency to reduce the effect of crosstalk. \label{fig:fdma_structures}}
\end{figure*}

The primary advantage of having a shared controller for all the unit cells is that only a single AC signal must be generated, which can be shared over multiple unit cells.
As the power is shared across multiple unit cells, a driver that dissipates more power can be implemented, while still maintaining a low dissipated power per qubit.
However, using a single shared driver requires each unit cell to be biased with a different magnetic field, as shown in Fig. \ref{fig:fdma_structures}(a), such that each electron spin is addressable with a different Larmor frequency, and it places additional requirements on the driving signal, which needs to drive one electron spin qubit without introducing crosstalk toward other electron spin qubits.
Different Larmor frequencies in each unit cell can be obtained by using a gradient in the permanent magnetic field or by locally generating an additional parallel DC field with a DC-current in a local coil.
Furthermore, using a shared driver requires the qubit controller to have a wider bandwidth since it needs to drive the different Larmor frequencies.
While this can be a significant source of power dissipation, the controller can be placed at a different temperature stage where more power can be dissipated, as only a single frequency-multiplexed cable needs to be routed towards another temperature stage \cite{VanDijk2019_jssc}. 
A more fundamental limitation is given by the nuclear spins that, even when the unit cells are biased with a different magnetic field, can still have overlapping Larmor frequencies due to their interaction with the electron spin, preventing direct driving and requiring the electron spin to perform operations on the nuclear spin (see Fig. \ref{fig:nv_system}).
Consequently, the number of nuclear spins which potentially can be addressed when directly driving the Larmor frequency is reduced.

While driving a single-qubit operation at a time via a shared frequency-multiplexed line may relax the driver requirements, it would be undesirable as it would significantly slow down the computation.
To avoid such a slowdown, the Rabi frequency could be increased for faster operations, but this would require larger currents in the coil and more frequency spacing between electron spin qubits to avoid infidelity, as seen by Eq. (\ref{eq:fdma_driving}).
Alternatively, multiple tones can be applied simultaneously to drive parallel operations, but this increases the peak and RMS currents, resulting in more Joule heating in a (non-superconductive) coil\footnote{Section \ref{sec:sys_impl} shows that \SI{100}{\milli \arms} can drive 20 qubits, but that already results in an excessive \SI{10}{\milli\watt} of dissipation in a \SI{1}{\ohm} coil}, which can heat up the color center and reduce its fidelity \cite{Sukachev2017}.
While superconducting coils can alleviate this issue, the driver must still reach higher output powers, higher peak currents and stricter linearity requirements, in addition to the superconducting coil requiring a higher critical current.

Scaling up the quantum processor involves both further adapting the magnetic-field gradient and increasing the controller bandwidth. Although this may be feasible up to tens and maybe hundreds of unit cells, it may become unfeasible beyond that.
More coils could then be introduced, where each coil drives a unit cell with a similar set of Larmor frequencies and is driven by a shared controller, resulting in a hybrid solution using multiple shared coils, as shown in Fig. \ref{fig:fdma_structures}(b).

\subsection{Dedicated driver}
Using an individual driver and coil per unit cell relieves the constraints on the number of nuclear-spin qubits that can be addressed without using the electron spin.
Furthermore, the driving signals for a unit cell are inherently attenuated at the neighboring cells thanks to the distance between the cells and the angle of the induced AC magnetic field, reducing the infidelity due to the crosstalk and allowing each color center to be tuned to the same Larmor frequency, as shown in Fig. \ref{fig:fdma_structures}(c).
This simplifies the design of the DC bias, as each unit cell ideally has the exact same DC magnetic field, which can be generated with a larger permanent magnet, at least for the major part (Fig. \ref{fig:nv_scalable}).
Only the inhomogeneities of the large permanent field must be compensated to ensure each unit cell has the same Larmor frequency.
If the infidelity due to crosstalk is too high in such architecture, e.g., because the unit cells are small, the Larmor frequency of neighboring cells can be detuned from each other to further reduce the infidelity, which would require additional local tuning range of the DC magnetic field in each unit cell (Fig. \ref{fig:fdma_structures}(d)).
Alternatively, the inhomogeneity can be left uncompensated if the bandwidth of the CMOS controller is large enough to compensate for the difference in Larmor frequency of the different unit cells.

\begin{figure}[t]
    \centering
    \includegraphics[width=0.95\linewidth]{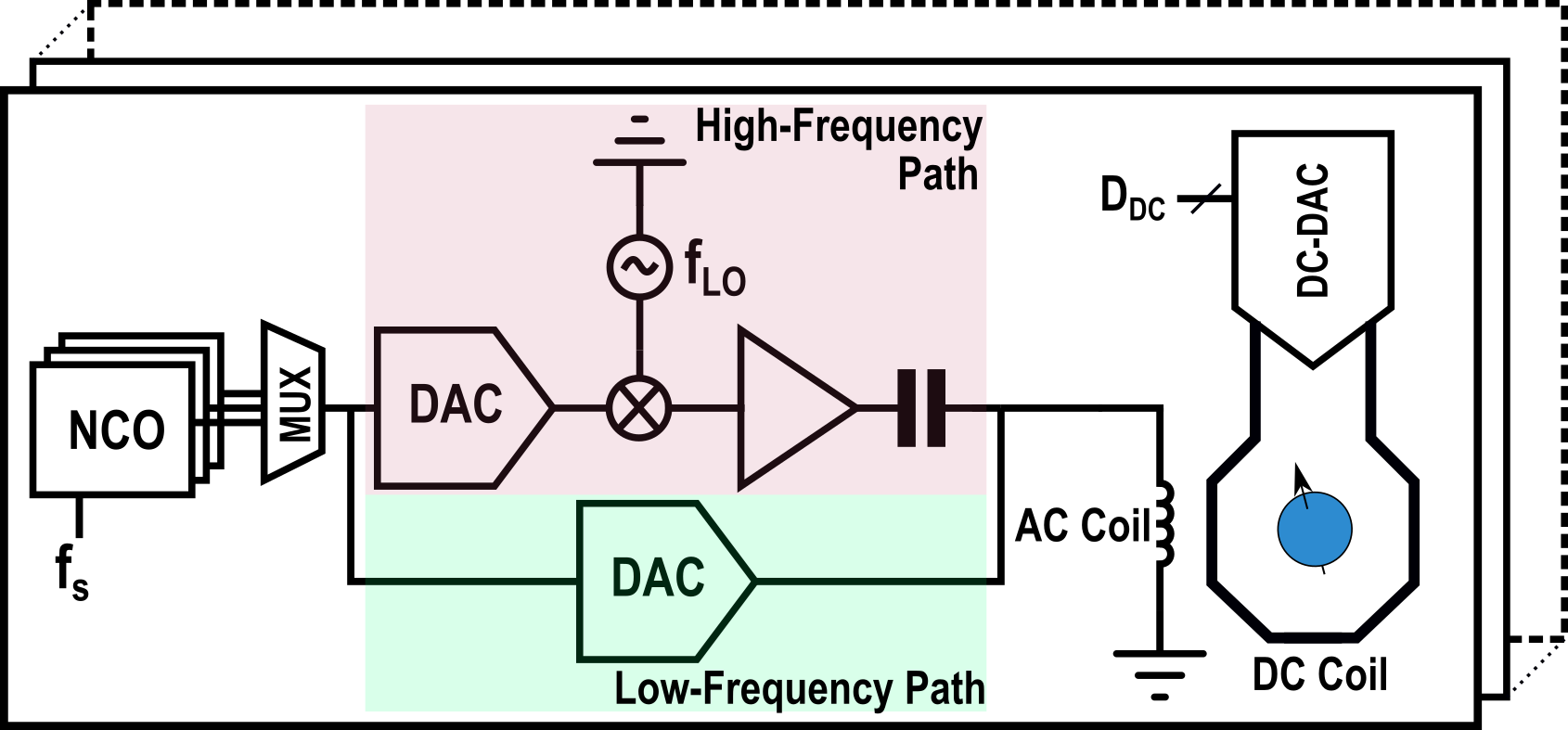}
    \caption{Generic qubit controller for single-qubit rotations based on prior work \cite{Frank2022, Park2021, VanDijk2020}. The electron spin is driven via the high-frequency path (red), while the nuclear spins are driven via the low-frequency path (green). The DC-magnetic-field generation can be used to compensate for the DC magnetic field inhomogeneity or for tuning the Larmor frequency, as in Fig.\ref{fig:fdma_structures}(d).}
    \label{fig:system_architecture}
\end{figure}

Each unit cell will require a dedicated controller and coils to locally generate the AC and DC signals for the qubits, enabling driving the coils directly, avoiding the typically employed 50-\SI{}{\ohm} impedance matching, and minimizing the current in the individual coils (unlike the shared-driver scenario), but also requiring more functionality at the unit-cell level compared to a shared driver and introducing additional power dissipation due to the required dedicated controller.
A generic block diagram of such a controller, inspired by prior work \cite{Frank2022, Park2021, VanDijk2020} is shown in Fig. \ref{fig:system_architecture}.
In this architecture, the NCO's track the Larmor frequencies of the electron and nuclear spins and synthesize the baseband waveforms that are converted to analog signals by the DACs.
In the high-frequency path for the electron spins, the signals are up-modulated, while this isn't required in the low-frequency path for the nuclear spins. 
Both paths are combined to drive a shared coil.
It is worthwhile reducing the power and area of the circuits of the components in Fig. \ref{fig:system_architecture}, as it improves scalability and relaxes the permanent-field inhomogeneity thanks to the smaller pitch of the unit cells.

\begin{figure}
    \centering
    \includegraphics[width=0.99\linewidth]{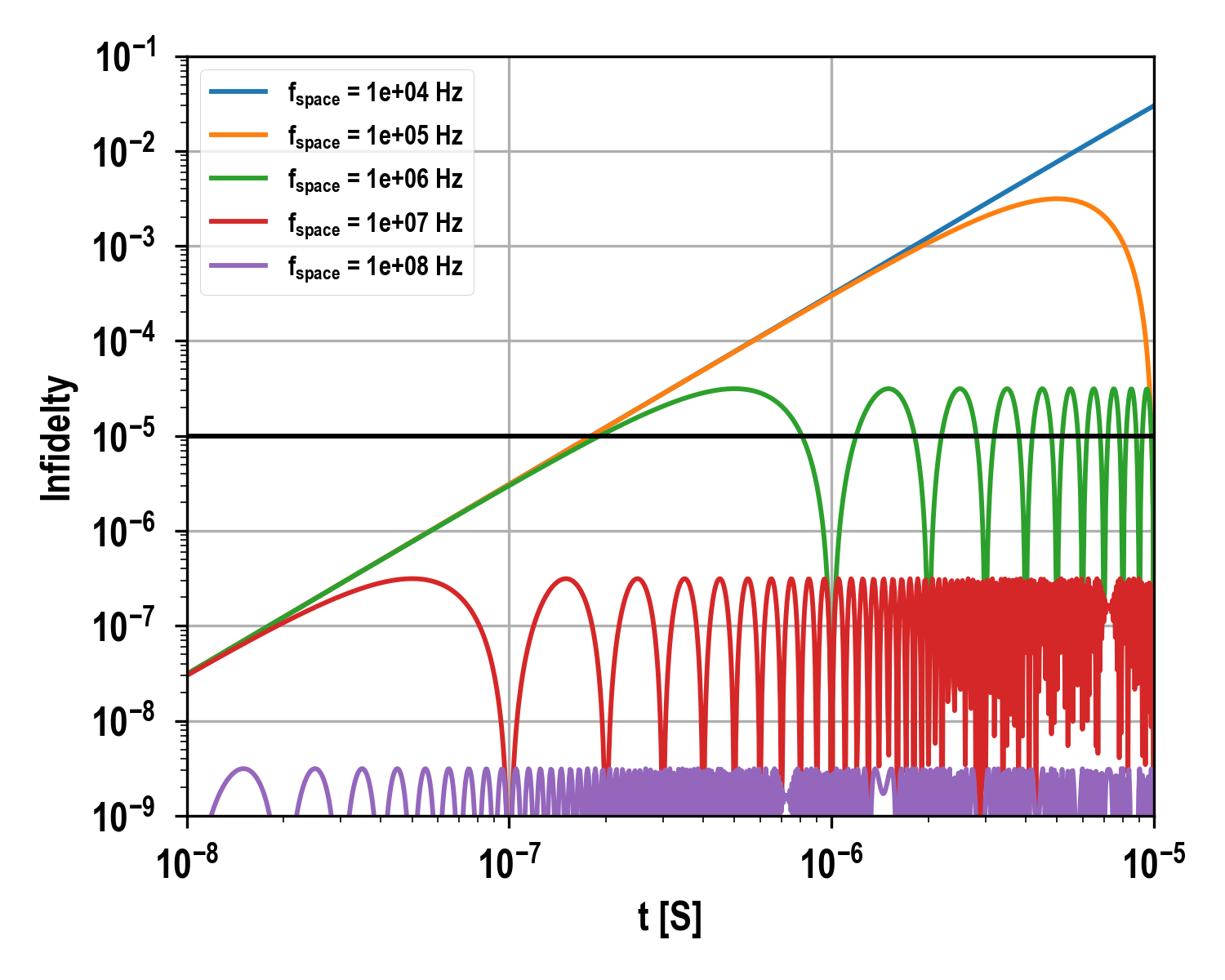}
    \caption{Infidelity due to crosstalk assuming a \SI{20}{\milli \ampere_{pk}} signal in the LO distribution, computed using eq. (\ref{eq:fdma_driving}) with $\theta=\omega_{R,addr} t$. 
   The coupling from the LO wire to the qubit location is simulated for a horizontal distance of \SI{500}{\micro\meter} (assuming routing the signal around a \SI{1}{\milli\meter}$\times$\SI{1}{\milli\meter} unit cell) and a height of \SI{15}{\micro\meter} between the wire and qubit due to the layer stack, resulting in a $\beta$ = \num{1.1e-3}. $\alpha$ varies with $f_{space}$. The black line indicates a target $10^{-5}$ infidelity.}
    \label{fig:LO_detuning}
\end{figure}

Each controller of Fig. \ref{fig:system_architecture} requires an LO signal, typically close to the Larmor frequency to limit the bandwidth of the baseband section.
Fortunately, since all qubits can be biased with a similar Larmor frequency, a single LO can be shared across the unit cells. 
However, a pervasive distribution network of such reference frequency is a potential source of interference, which can lower the fidelity of the qubit.
Active compensation of the LO leakage can be introduced but this may consume additional power and reduce the output range of the driver, while also requiring additional calibration \cite{yoo2023design}.
Alternatively, an LO frequency needs to be selected such that it is sufficiently detuned from all electron spins and thus does not contribute too much infidelity.
Fig. \ref{fig:LO_detuning} shows that a negligible impact on fidelity can be obtained in typically expected conditions with a detuning of just \SI{10}{\mega \hertz}, which would not significantly impact the power dissipation of the driver.

Compared to the shared driver, dedicated drivers have the advantage that the number of nuclear spin qubits that can be addressed is increased.
Furthermore, considering the scalability of the quantum processor, using separate unit cells is favorable since adding more unit cells requires only compensating for a larger inhomogeneity, which should not increase the power per unit cell significantly.
The open challenge is however that the controller in each unit cell must offer a power-efficient implementation, as the total dissipated power will scale directly with the number of unit cells.
Because of those advantages and the shortcomings of the shared driver, a separate driver implementation is investigated in the following to investigate the feasibility of generating the AC and DC magnetic fields with a low power and sufficient fidelity.


\section{Power estimation}
\label{sec:sys_impl}
\subsection{Coil design}

\begin{figure}[t]
    \centering
    \includegraphics[width=0.99\linewidth]{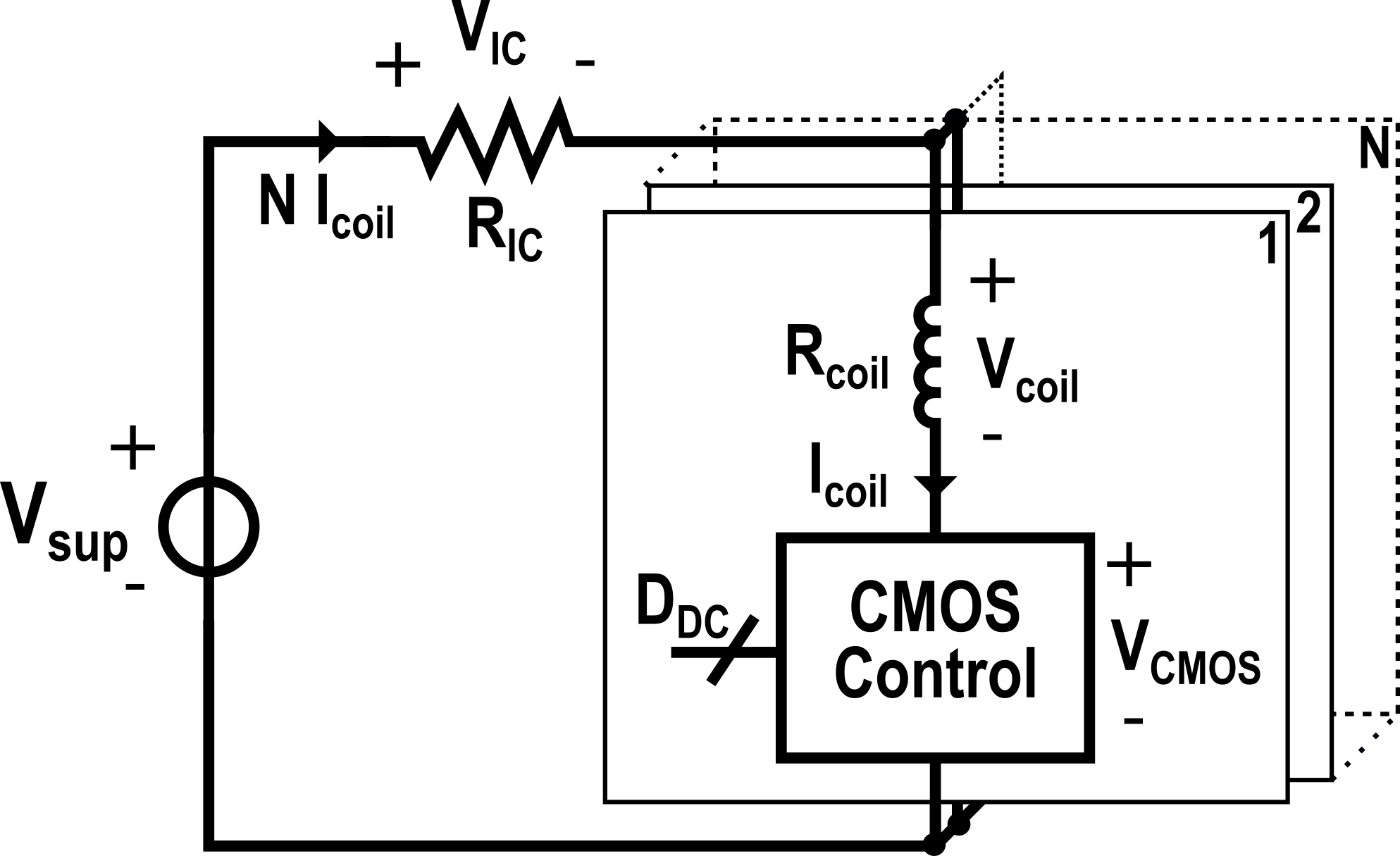}
    \caption{Simplified electrical circuit for analysis of the power dissipation. Each unit cell has a coil and a CMOS controller while the interconnect is shared across N unit cells. The CMOS controller could also be placed on the other side of the coil to allow reversing the current polarity but this is not shown for simplicity.}
    \label{fig:circuit_DC}
\end{figure}

The most effective way to apply the large DC bias magnetic field to all the unit cells is by using a global field, for instance, generated externally by a permanent magnet, as illustrated in Fig. \ref{fig:nv_scalable}, since it would not dissipate power close to the cooled-down qubits.
However, depending on the controller implementation, locally generated magnetic fields may still be required to compensate for magnetic field inhomogeneity.
By simulating with COMSOL\cite{pryor2009multiphysics} a global Helmholtz coil\footnote{Simulated with radius of \SI{50}{\milli\meter}, coil width and thickness of \SI{50}{\milli\meter}, and a coil spacing of \SI{40}{\milli\meter}.} over a chip area of \SI{10}{\milli\meter} $\times$ \SI{10}{\milli\meter}, the inhomogeneity to be compensated amounts to \SI{\pm 2.4}{\gauss} for a bias field of \SI{2000}{\gauss}.
Compensation of such an inhomogeneity can be achieved via permanent micro-magnets in each unit cell, but their accurate tuning is technologically very challenging.
Instead, by running a current through a coil close to the color center, the DC magnetic field can be precisely controlled, requiring electronics that regulates the current through the coil, as shown in Fig. \ref{fig:circuit_DC}.
Alternatively, a superconducting loop can be biased by a continuously recirculating permanent current \cite{Lacy2020} to avoid any power dissipation.
Nevertheless, this approach would introduce extra complexity in fabrication and integration, in addition to other unknown risks, such as the stability of these DC fields in the presence of the AC fields for qubit operations.
For simplicity, it is assumed that the current needs to be actively driven through a conductive coil, such that no superconducting effects (i.e., Meissner effect) should be taken into account.

\begin{figure*}
    \centering
    \includegraphics[width=0.99\textwidth]{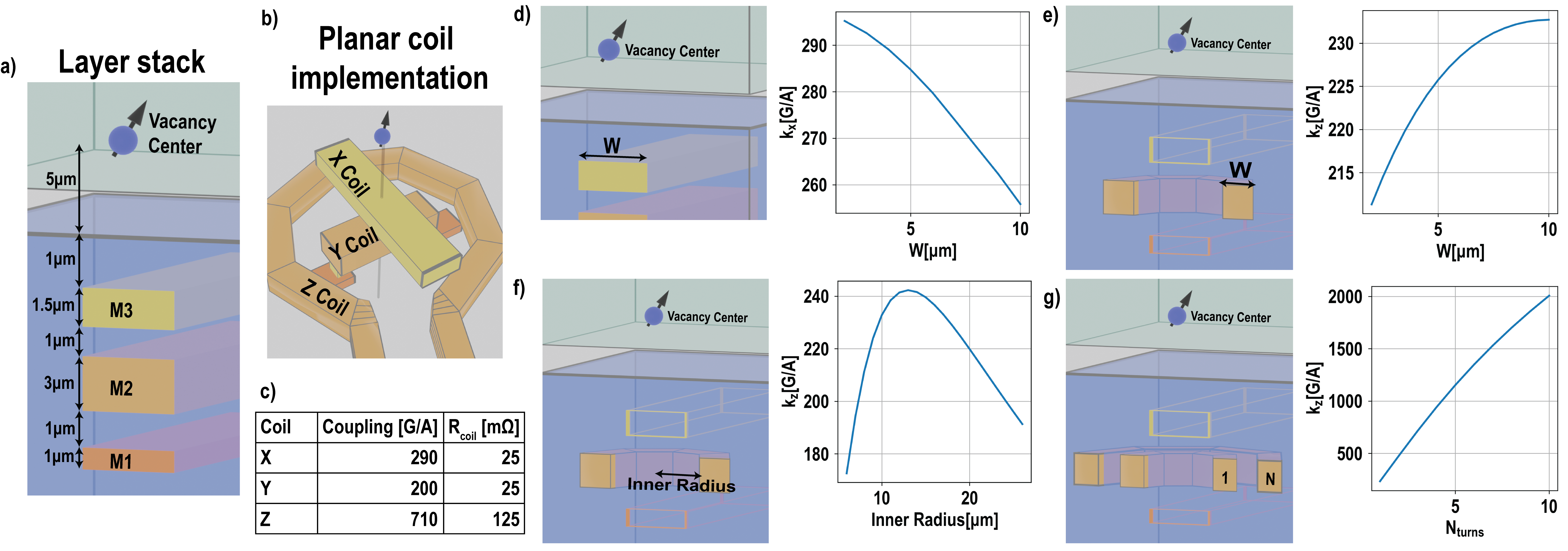}
    \caption{Magnetic field simulations to derive the coupling $k$ for the different coils: a) Metal stack in the cryo-CMOS chip for coil implementation; b) Planar implementation of the $x, y, z$-coils; c) Simulated coupling and estimated resistance at \SI{4}{\kelvin} for the x,y,z-coils based on the metal stack d) Coupling $k_x$ as a function of the line width. $k_y$ is simulated similar to $k_x$, but uses a the lower metal leading to reduced coupling. Note that the metal layer of the $x$ and $y$-coils can be swapped. e) Coupling $k_z$ versus the line width of a single turn z-coil with an inner radius of \SI{8}{\micro\meter}; f) Coupling $k_z$ versus the inner radius of a single turn Z-coil for a line width of \SI{2}{\micro\meter}; g) Coupling $k_z$ versus the number of turns with a width of \SI{2}{\micro\meter}, pitch of \SI{3}{\micro\meter} and inner radius of \SI{10}{\micro\meter}.\label{fig:coil_sweeps}}
\end{figure*}

To understand the constraints in power dissipation introduced by the electronic interface, the DC and AC requirements in Table \ref{tab:specifications}, which is expressed in terms of magnetic field, must be converted into coil driving currents by analyzing the coil current-to-magnetic-field coupling $k_{x,y,z}$ in \SI{}{\gauss/\ampere} together with the coil resistance $R_{coil}$ in \SI{}{\ohm}.
Here, the $x,y,z$ subscript indicates the axis of the applied magnetic field, where $z$ is aligned with $B_\parallel$, while $x$ and $y$ are in the $B_\perp$ orientation.
Ideally, the coefficients $k_{x,y,z}$ should be maximized while having a low coil resistance to minimize Joule heating, so as to obtain a low power dissipation for a low driving current.
However, even if the coil resistance becomes very low, other sources of dissipation in the system can dominate, for instance due to the CMOS control and interconnect, as illustrated in Fig. \ref{fig:circuit_DC}.
Since many parameters affect $k$ and $R_{coil}$, a generic CMOS metal stack is used to estimate distances, optimize the coupling and extract resistances as shown in Fig. \ref{fig:coil_sweeps} and leading to the parameters in Fig.\ref{fig:coil_sweeps}(c).
The coupling parameters $k$ for the different coils allow translating Table \ref{tab:specifications} to specifications in the electrical domain, indicating that peak AC currents of \SI{8.3}{\milli \ampere_{pk}} and \SI{15.7}{\milli \ampere_{pk}} are needed to achieve a Rabi frequency of \SI{5}{\mega\hertz} and \SI{5}{\kilo\hertz} for the electron and nuclear spins, respectively (assuming that both drive the same coil) and DC currents up to \SI{\pm3.4}{\milli\ampere} are needed to correct the \SI{\pm 2.4}{\gauss} magnetic field inhomogeneity introduced by the permanent magnet.


\subsection{Control Electronics} \label{ssec:cont_elec}
With an indication of the coupling and required current levels, a more detailed look is cast on the different blocks shown in Fig. \ref{fig:system_architecture} and Fig. \ref{fig:circuit_DC} so that their power consumption can be estimated, leading towards their optimization presented in Section \ref{subs:power_opt}.

\subsubsection{DC Magnetic Field Generator}
From Fig. \ref{fig:circuit_DC}, the power dissipation of the DC control per unit cell can be expressed as
\begin{align}\label{eq:p_mfg_dc}
    P_{DC}&= \frac{(NI_{coil})^2R_{IC}}{N}+I_{coil}^2(R_{on}+  R_{coil}) + P_{cir} \nonumber \\
            &= I_{coil}^2(N R_{IC} + R_{on}+ R_{coil}) + P_{cir}
\end{align}
where $I_{coil}$ is the maximum current required in the coil, $R_{on}$, $R_{coil}$ and $R_{IC}$ are the on-resistance of the CMOS control, coil resistance and interconnect resistance, respectively, $N$ is the number of unit cells, and $P_{cir}$ is the auxiliary power required for the regulation loop.
For the circuit, an operating temperature of \SI{4}{\kelvin} is assumed, reducing the different resistances $R_{on}$ (2$\times$), $R_{coil}$ (4$\times$) and $R_{IC}$ (4$\times$) with respect to RT \cite{patra2017cryo,Patra2018}.
Simulations in a commercial CMOS process verify that a transistor in triode can achieve a $R_{on}\approx$\SI{0.25}{\ohm} at \SI{4}{\kelvin} when occupying an area of \SI{2500}{\micro\meter^2}. 
$R_{on}$ can be reduced further, but at the expense of noise performance, since thermal noise from the CMOS circuit is proportional to $\frac{4kT}{R}$, and area, as a low resistance requires larger transistors, whereas a larger $R_{on}$ results in more dissipation, effectively trading power dissipation for infidelity in Table \ref{tab:specifications}.

The interconnect resistance $R_{IC}$ is estimated to be around \SI{12.5}{\milli \ohm} at \SI{4}{\kelvin}.
\footnote{$R_{IC}$ has been extracted for a \SI{10}{\milli\meter} $\times$ \SI{10}{\milli\meter} chip taking the power grid resistance from an unit cell in the chip center to the pads on the chips periphery. The chips periphery allows for more than 700 staggered bondpads to further reduce the interconnect resistance to the PCB.}
The auxiliary circuit power $P_{cir}$ is estimated to be around \SI{100}{\micro \watt} since some form of control is needed that accurately sets the current, but the circuit implementation is left as future work.
While the exact value of $I_{coil}$ that each unit cell needs depends on the 3D-integration and the placement of the quantum processor in the bias field, a pessimistic assumption can be made that each unit cell will require the maximum current $I_{coil}$, such that the power dissipation required for DC biasing can be estimated.

\subsubsection{Digital Section of the AC driver}
The clock frequency used in the baseband of the AC driver directly affects the power consumption. 
When using FDMA as in \cite{VanDijk2020}, high-bandwidth DACs are required, thus asking for a high clock frequency and increasing the power dissipation in the NCOs and DACs (Fig. \ref{fig:system_architecture}).
As discussed in Section \ref{sec:sys_arch}, the qubit controller of this work targets a single electron spin and 9 nuclear spins (Fig. \ref{fig:system_architecture}) and thus can have a much lower bandwidth, allowing the power dissipation to be significantly reduced.
For the electron spin, the bandwidth of the DAC needs to be large enough to space the electron spin sufficiently from the LO ($f_{space,LO}$) to prevent any accidental driving from clock lines to the qubit. 
Furthermore, a larger bandwidth is needed if inhomogeneities of the Larmor frequencies are compensated in the frequency domain ($f_{comp}$) to avoid local DC biasing.
In the rest of the Section, it is assumed that the clock frequency of the controller needs to fulfill: 
\begin{equation}
f_s \geq 2.5\times \max[f_{space,LO}, f_{comp}]
\end{equation}
such that sufficient bandwidth is available in the controller. This work assumes $f_{space,LO}$= \SI{10}{\mega\hertz} and $f_{comp}$ between \SI{0}{\mega\hertz} and \SI{15}{\mega\hertz}.
For the nuclear spins, the bandwidth of the DAC needs to be sufficiently large in order to synthesize the signals in the frequency range of Table \ref{tab:specifications} ($f_{0,c}$).
This part of the controller could typically operate with a lower sampling frequency of $2.5\times f_{0,c}$, but depending on $f_{space,LO}$ and $f_{comp}$, the difference in clock frequency may be small and may complicate the controller design with multiple clock domains.
Therefore, here it is assumed that the same clock frequency is used everywhere in the AC driver.

Each unit cell will require at least 10 NCOs, 1 for the electron spin and 9 for the nuclear spins.
As one can observe in Table \ref{tab:specifications}, the required frequency accuracy for the various spins differs, meaning that the NCO's for the electron and nuclear spin will require different number of bits.
The frequency resolution $\Delta f$ for a given number of bits $N_{bits}$ in the NCO is $\Delta f = \frac{f_s}{2^{N_{bits}+1}}$ \cite{VanDijk2020}.
Together with eq. (\ref{eq:inf_df}), $N_{bits}$ can be computed for a required fidelity: 
\begin{equation}
    N_{bits} = \Biggl \lceil{\log_2 \Biggl(\frac{\pi f_s T_{op,e/n}}{\acos\bigl(\sqrt{F}\bigr)}\Biggr) - 1} \Biggr \rceil
\end{equation}
where $T_{op,e/n}$ is the operation time of the nuclear/electron spin and $f_s$ the clock frequency of the NCO's \cite{VanDijk2020}.
To achieve \num{1e-5} infidelity\footnote{The NCO introduces only a part of the frequency inaccuracy, hence it needs to contribute low infidelity.} with $f_s$ = \SI{25}{\mega \hertz}, $T_{op,e}$ = \SI{100}{\nano \second}, and $T_{op,n}$ = \SI{100}{\micro \second} (see Table \ref{tab:specifications}), 11 bits are needed in the NCO to track the electron spin frequency and 21 bits are needed to track the nuclear spin frequency \cite{VanDijk2020} assuming both use the same $f_s$.
The power dissipation of the NCO can then be calculated with
\begin{equation}\label{eq:nco_pdiss}
    P_{nco} = E_{bit} f_s \Biggl \lceil{\log_2\Biggl(\frac{\pi f_s T_{op,e/n}}{\acos\bigl(\sqrt{F}\bigr)}\Biggr) - 1} \Biggr \rceil,
\end{equation}
where $E_{bit}$ is the energy per bit in the NCO.
By using $E_{bit}=$\SI{84}{\femto \joule/\bit} from \cite{VanDijk2020} for a \SI{22}{\nano\meter} technology, this results in a power dissipation of \SI{46}{\micro\watt} and \SI{88}{\micro\watt} for the NCOs of the electron and nuclear spin, respectively, leading to a total power dissipation of \SI{0.84}{\milli\watt} for all NCOs combined.
Depending on the full controller implementation, other digital blocks that operate at $f_s$ may be needed for calibration or pulse shaping.
Nevertheless, it is expected that the NCOs will consume most power and hence power estimations in subsection \ref{subs:power_opt} will only consider the NCOs \cite{VanDijk2020}.

\subsubsection{Analog circuits in the AC driver}
The analog circuits of the qubit control system consist of the DACs, mixer and output amplifier. 
The DACs and mixer may have some dependence on the clock frequency of the digital, and will add additional power dissipation when enabled.
However, the largest source of dissipation is expected to come from the output amplifier, since it drives large peak currents to the coil.
Hence, this section zooms in on the amplifier to provide an estimate of its dissipation.

The power dissipated by the output amplifier for driving operations on a NV center electron spin can be given by
\begin{equation}\label{eq:p_electron_operation}
    P_{rms}^{NVe,op} = \frac{f_{r,e}}{\gamma_e k_{coil}}V_{DD}D_{e,op}
\end{equation}
which is derived from eq. (\ref{eq:Rabi}), converting the magnetic field to a current with $k_{coil}$, and considering a supply voltage $V_{DD}$.
A duty-cycle factor $D_{e,op}$ is added here, since the output amplifier is not continuously active, and often there are long delays between electron operations as shown in Fig. \ref{fig:nv_system}(d) with the effective on-time of the amplifier, $D_{e,op}$, being less than 10\%.
For a Rabi frequency of \SI{5}{\mega \hertz} and $k_{coil}$ = \SI{290}{\gauss \per \ampere}, a peak current of \SI{8.3}{\milli\ampere_{pk}} is required\footnote{Here, $k_{coil}$ is $k_x$ from Fig. \ref{fig:coil_sweeps}.}.
When assuming a supply of \SI{1.1}{\volt}, this leads to an effective power dissipation of \SI{6.5}{\milli\watt_{rms}} when continuously driving an operation and assuming \SI{100}{\%} amplifier efficiency.
With a duty cycle of \SI{10}{\%}, this leads to an effective dissipation of \SI{650}{\micro \watt} if no static dissipation is present.

For nuclear spin operations, the power dissipation of the output stage can be estimated by
\begin{equation}\label{eq:p_carbon_operation}
    P_{rms}^{c,op} = \frac{f_{r,c}}{\sqrt{2}\gamma_c k_{coil}}V_{sup}.
\end{equation}
While largely similar to eq. (\ref{eq:p_electron_operation}), a factor $\frac{1}{\sqrt{2}}$ is introduced and a different supply is assumed here.
For the target Rabi frequency of \SI{5}{\kilo\hertz}, a peak current of \SI{16.1}{\milli\ampere_{pk}} will be required.
Compared to the electron spin operations, the nuclear spin operations are long and are driven almost continuously, so the effective power dissipated is not reduced by having a low duty-cycle.
Hence, to limit the power dissipated for nuclear spin operations only one degree of freedom is available. 
Similar to the DC magnetic field generator, one needs to reduce the supply $V_{sup}$ to reduce the power dissipated.

Assuming $V_{sup}$=\SI{100}{\milli \volt} results in a power dissipation of \SI{1.1}{\milli\watt_{rms}}.
To allow for the lower supply $V_{sup}$, separate output stages for both electron and nuclear spin operations are required, eventually driving the same coil as shown in Fig. \ref{fig:system_architecture}.

\subsection{Power Estimation}
\label{subs:power_opt}
Combining eq. (\ref{eq:p_mfg_dc}), (\ref{eq:nco_pdiss}), (\ref{eq:p_electron_operation}) and (\ref{eq:p_carbon_operation}), the power dissipation for a unit cell can be estimated to analyze dominant sources of power dissipation.
The parameters that are listed in subsection \ref{ssec:cont_elec} are used, providing values for the AC signal generation, the sampling frequency and the DC magnetic field generation.
However, for the dissipation of the DC magnetic field and the sampling frequency of the qubit controller, a trade-off is present, since the inhomogeneity of the magnetic field can be compensated by tuning the current through a local DC coil (eq. (\ref{eq:p_mfg_dc})) or by increasing $f_s$ of the digital (eq. (\ref{eq:nco_pdiss})).
Here, two cases are considered:
In the first case, the magnetic field inhomogeneity is fully compensated by a local magnetic field, resulting in a fixed $f_s$.
In the second case, the inhomogeneity of the magnetic field is compensated by increasing the bandwidth of the DAC, so no DC magnetic field correction is performed but $f_s$ increases with the inhomogeneity.

\begin{figure}[t]
    \centering
    \includegraphics[width=0.99\linewidth]{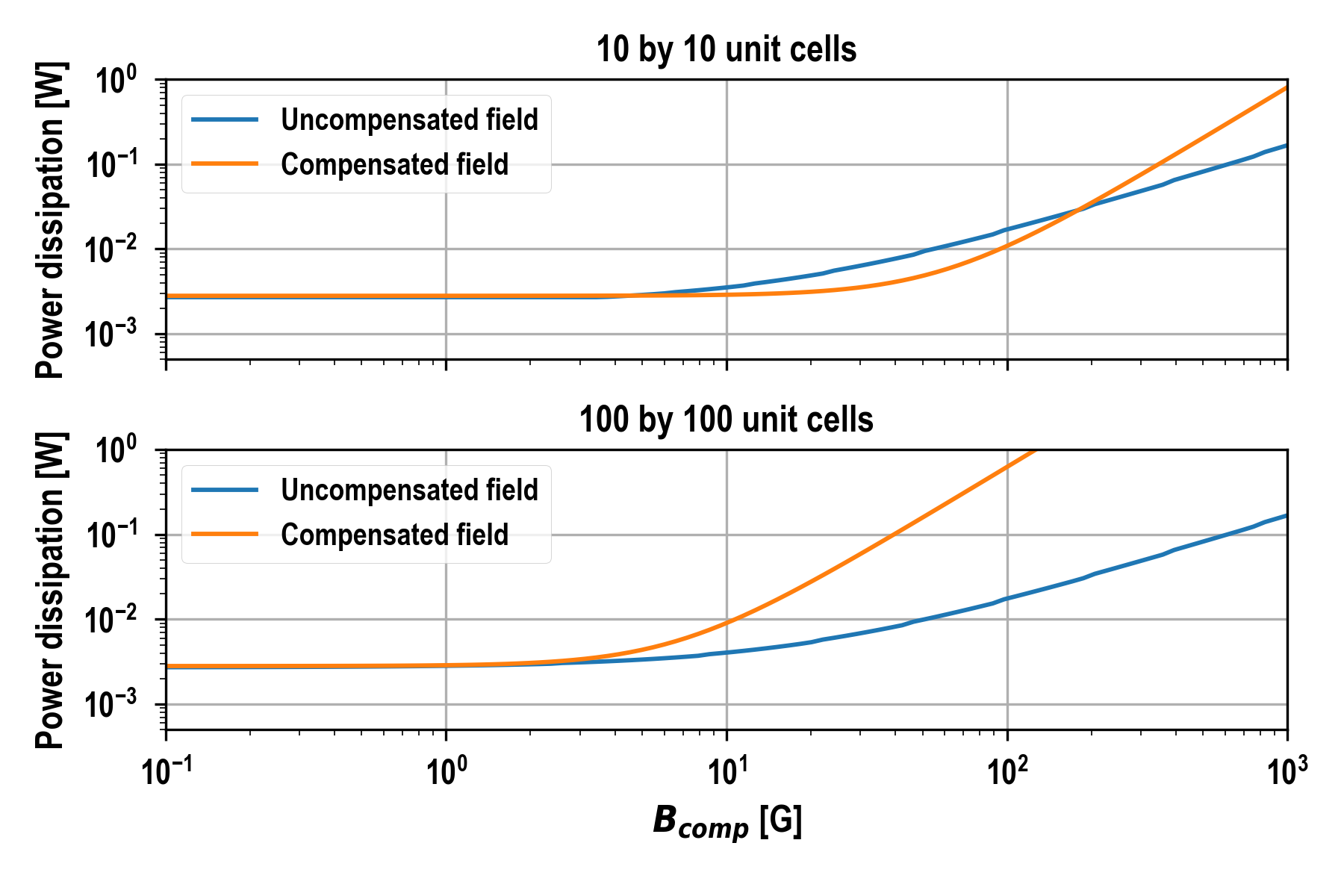}
    \caption{Comparison of the power dissipated per unit cell at \SI{4}{\kelvin} by compensating the DC magnetic field by changing the DC magnetic field locally or by increasing the clock speed of the digital. The power is plotted versus to-be-compensated magnetic field for 10 by 10 unit cells (top) and 100 by 100 unit cells (bottom).}
    \label{fig:power_tradeoff}
\end{figure}

The power per unit cell for the two cases is plotted in Fig. \ref{fig:power_tradeoff}, since hybrid inhomogeneity compensation typically lies between these two cases.
Here, the power dissipation is computed for both 10 by 10 and 100 by 100 unit cells by increasing the number of unit cells $N$, such that the influence of more unit cells on the power dissipation becomes clear.
At a low inhomogeneity of the bias field, the power dissipation is primarily dominated by the AC signal generation and the $f_{space}$ requirement for the qubit controller, while $P_{cir}$ for the DC magnetic field generator adds negligible power dissipation.
When a large magnetic-field inhomogeneity needs to be compensated, the power dissipated in the DC magnetic field generator is larger due to the scaling of the power, i.e., $P \propto |B_{comp}^2|$, while the controller scales $P \propto |B_{comp}|$.
There is however an intermediate region for 10 x 10 unit cells where the power dissipation is lower when compensating the DC magnetic field.
In case the magnetic field is compensated using DC coils, the power dissipated in the interconnect will increase significantly.
Consequently, unless the power dissipation in the interconnect can be reduced, frequency-domain compensation as illustrated in Fig. \ref{fig:power_tradeoff} is more power efficient for larger quantum processors.

While some of the assumed parameters for Fig. \ref{fig:power_tradeoff} might change, the methodology and reasoning can be re-used for future analysis of quantum processors.
Simulation of Helmholtz coils that can generate the permanent magnetic field indicate that inhomogeneities in the order of \SI{\approx0.25}{\percent} can be expected, resulting in an inhomogeneity of \SI{\pm2.4}{\gauss} and power dissipation of \SI{3}{\milli\watt} for \SI{2000}{\gauss}.
In practice, better permanent biasing coils can be engineered, but also integration imperfections can require compensation, meaning that some form of compensation may always be required in a similar field range.
Based on the estimations, existing dilution refrigerator could host more than 100 unit cells at \SI{1}{\kelvin}, which would integrate more than 1000 qubits together with their required control electronics.

The analysis further indicates focus points for the quantum processor. For instance, with low inhomogeneity of the magnetic field, the power dissipation will be dominated by the AC signal generation for the electron and nuclear spins, meaning it is desired to reduce the current and power needed to perform qubit rotations.
While the current can be reduced by improving the coupling of the coil, additional research in efficiently generating high currents is also required.
For larger inhomogeneities, frequency compensation in the digital domain is favorable thanks to the linear scaling and since the power dissipation per unit cell is independent of the number of unit cells present.
Nevertheless, if lower interconnect resistances can be realized, e.g., through backside power delivery \cite{hossen2019power}, compensating the DC magnetic field can become more attractive for larger quantum processors.
Furthermore, it should be investigated if the DC magnetic field generator and AC controller can actually achieve the estimated low power operation without affecting the qubit fidelity, thus requiring additional prototyping and validation with qubits.

\section{Conclusions}
\label{sec:conclusion}
This paper presents a high-level analysis on creating a 3D-integrated quantum processor using color centers in diamond and ranges from introducing and understanding the qubits in the system, to deriving their requirements, finding a suitable system architecture and estimating the power dissipation.
In the quantum processor, identical unit cells are combined, and each unit cell requires full functionality to operate the qubits, such as DC biasing, electrical AC signals, and optical signals.
Here, a system architecture that uses separate drivers and coils in each unit cell is favored as it maximizes the available number of qubits. 
Consequently, the electronics footprint will require physical spacing between the qubits, thus inherently reducing the crosstalk and not needing frequency multiplexing. 
However, a full controller that generates AC and DC signals is needed in each unit cell, which requires a low-power implementation to maximize scalability.
Estimations indicate that such controller can be implemented in a low-power manner, allowing for more than 100 unit cells to operate with less than \SI{1}{\watt} of power dissipation. 
While this work demonstrates how architectural approaches can be beneficial for reducing the power dissipation in intermediate- and large-scale vacancy-center quantum processors, the proposed methodology can be applied to similar qubit technologies to support the system engineering of future large-scale quantum processors.

\section{Acknowledgment}

We gratefully acknowledge support from the collaboration between Fujitsu Limited and Delft University of Technology, co-funded by the Netherlands Enterprise Agency under project number PPS2007. Data repository in \cite{dataset}.

\bibliographystyle{IEEEtran}
\bibliography{bibliography.bib}

\begin{thebibliography}{10}
\providecommand{\url}[1]{#1}
\csname url@samestyle\endcsname
\providecommand{\newblock}{\relax}
\providecommand{\bibinfo}[2]{#2}
\providecommand{\BIBentrySTDinterwordspacing}{\spaceskip=0pt\relax}
\providecommand{\BIBentryALTinterwordstretchfactor}{4}
\providecommand{\BIBentryALTinterwordspacing}{\spaceskip=\fontdimen2\font plus
\BIBentryALTinterwordstretchfactor\fontdimen3\font minus \fontdimen4\font\relax}
\providecommand{\BIBforeignlanguage}[2]{{%
\expandafter\ifx\csname l@#1\endcsname\relax
\typeout{** WARNING: IEEEtran.bst: No hyphenation pattern has been}%
\typeout{** loaded for the language `#1'. Using the pattern for}%
\typeout{** the default language instead.}%
\else
\language=\csname l@#1\endcsname
\fi
#2}}
\providecommand{\BIBdecl}{\relax}
\BIBdecl

\bibitem{feynman2018simulating}
R.~P. Feynman \emph{et~al.}, ``Simulating physics with computers,'' \emph{Int. j. Theor. phys}, vol.~21, no. 6/7, 2018.

\bibitem{cao2018potential}
Y.~Cao, J.~Romero, and A.~Aspuru-Guzik, ``Potential of quantum computing for drug discovery,'' \emph{IBM Journal of Research and Development}, vol.~62, no.~6, pp. 6--1, 2018.

\bibitem{cao2019quantum}
Y.~Cao, J.~Romero, J.~P. Olson, M.~Degroote, P.~D. Johnson, M.~Kieferov{\'a}, I.~D. Kivlichan, T.~Menke, B.~Peropadre, N.~P. Sawaya \emph{et~al.}, ``Quantum chemistry in the age of quantum computing,'' \emph{Chemical reviews}, vol. 119, no.~19, pp. 10\,856--10\,915, 2019.

\bibitem{fowler2012surface}
A.~G. Fowler, M.~Mariantoni, J.~M. Martinis, and A.~N. Cleland, ``Surface codes: Towards practical large-scale quantum computation,'' \emph{Physical Review A}, vol.~86, no.~3, p. 032324, 2012.

\bibitem{Patra2018}
B.~Patra, R.~M. Incandela, J.~P.~V. Dijk, H.~A. Homulle, L.~Song, M.~Shahmohammadi, R.~B. Staszewski, A.~Vladimirescu, M.~Babaie, F.~Sebastiano, and E.~Charbon, ``Cryo-cmos circuits and systems for quantum computing applications,'' \emph{IEEE Journal of Solid-State Circuits}, vol.~53, pp. 309--321, 1 2018.

\bibitem{Arute2019}
F.~Arute, K.~Arya, R.~Babbush, D.~Bacon, J.~C. Bardin, R.~Barends, R.~Biswas, S.~Boixo, F.~G. Brandao, D.~A. Buell, B.~Burkett, Y.~Chen, Z.~Chen, B.~Chiaro, R.~Collins, W.~Courtney, A.~Dunsworth, E.~Farhi, B.~Foxen, A.~Fowler, C.~Gidney, M.~Giustina, R.~Graff, K.~Guerin, S.~Habegger, M.~P. Harrigan, M.~J. Hartmann, A.~Ho, M.~Hoffmann, T.~Huang, T.~S. Humble, S.~V. Isakov, E.~Jeffrey, Z.~Jiang, D.~Kafri, K.~Kechedzhi, J.~Kelly, P.~V. Klimov, S.~Knysh, A.~Korotkov, F.~Kostritsa, D.~Landhuis, M.~Lindmark, E.~Lucero, D.~Lyakh, S.~Mandrà, J.~R. McClean, M.~McEwen, A.~Megrant, X.~Mi, K.~Michielsen, M.~Mohseni, J.~Mutus, O.~Naaman, M.~Neeley, C.~Neill, M.~Y. Niu, E.~Ostby, A.~Petukhov, J.~C. Platt, C.~Quintana, E.~G. Rieffel, P.~Roushan, N.~C. Rubin, D.~Sank, K.~J. Satzinger, V.~Smelyanskiy, K.~J. Sung, M.~D. Trevithick, A.~Vainsencher, B.~Villalonga, T.~White, Z.~J. Yao, P.~Yeh, A.~Zalcman, H.~Neven, and J.~M. Martinis, ``Quantum supremacy using a programmable superconducting processor,'' \emph{Nature}, vol. 574,
  pp. 505--510, 10 2019.

\bibitem{alberts2021accelerating}
G.~J. Alberts, M.~A. Rol, T.~Last, B.~W. Broer, C.~C. Bultink, M.~S. Rijlaarsdam, and A.~E. Van~Hauwermeiren, ``Accelerating quantum computer developments,'' \emph{EPJ Quantum Technology}, vol.~8, no.~1, p.~18, 2021.

\bibitem{bardin2022beyond}
J.~Bardin, ``Beyond-classical computing using superconducting quantum processors,'' in \emph{2022 IEEE International Solid-State Circuits Conference (ISSCC)}, vol.~65.\hskip 1em plus 0.5em minus 0.4em\relax IEEE, 2022, pp. 422--424.

\bibitem{zettles202226}
G.~Zettles, S.~Willenborg, B.~R. Johnson, A.~Wack, and B.~Allison, ``26.2 design considerations for superconducting quantum systems,'' in \emph{2022 IEEE International Solid-State Circuits Conference (ISSCC)}, vol.~65.\hskip 1em plus 0.5em minus 0.4em\relax IEEE, 2022, pp. 1--3.

\bibitem{yoo2023design}
J.~Yoo, Z.~Chen, F.~Arute, S.~Montazeri, M.~Szalay, C.~Erickson, E.~Jeffrey, R.~Fatemi, M.~Giustina, M.~Ansmann \emph{et~al.}, ``Design and characterization of a $<$4-mw/qubit 28-nm cryo-cmos integrated circuit for full control of a superconducting quantum processor unit cell,'' \emph{IEEE Journal of Solid-State Circuits}, 2023.

\bibitem{Frank2022}
D.~J. Frank, S.~Chakraborty, K.~Tien, P.~Rosno, T.~Fox, M.~Yeck, J.~A. Glick, R.~Robertazzi, R.~Richetta, J.~F. Bulzacchelli, D.~Ramirez, D.~Yilma, A.~Davies, R.~V. Joshi, S.~D. Chambers, S.~Lekuch, K.~Inoue, D.~Underwood, D.~Wisnieff, C.~Baks, D.~Bethune, J.~Timmerwilke, B.~R. Johnson, B.~P. Gaucher, and D.~J. Friedman, ``A cryo-cmos low-power semi-autonomous qubit state controller in 14nm finfet technology,'' vol. 2022-February.\hskip 1em plus 0.5em minus 0.4em\relax Institute of Electrical and Electronics Engineers Inc., 2022, pp. 360--362.

\bibitem{Park2021}
J.~Park, S.~Subramanian, L.~Lampert, T.~Mladenov, I.~Klotchkov, D.~J. Kurian, E.~Juarez-Hernandez, B.~P. Esparza, S.~R. Kale, K.~T.~A. Beevi, S.~P. Premaratne, T.~F. Watson, S.~Suzuki, M.~Rahman, J.~B. Timbadiya, S.~Soni, and S.~Pellerano, ``A fully integrated cryo-cmos soc for state manipulation, readout, and high-speed gate pulsing of spin qubits,'' \emph{IEEE Journal of Solid-State Circuits}, vol.~56, pp. 3289--3306, 11 2021.

\bibitem{VanDijk2019_jssc}
J.~P. G.~V. DIjk, B.~Patra, S.~Subramanian, X.~Xue, N.~Samkharadze, A.~Corna, C.~Jeon, F.~Sheikh, E.~Juarez-Hernandez, B.~P. Esparza, H.~Rampurawala, B.~R. Carlton, S.~Ravikumar, C.~Nieva, S.~Kim, H.~J. Lee, A.~Sammak, G.~Scappucci, M.~Veldhorst, L.~M. Vandersypen, E.~Charbon, S.~Pellerano, M.~Babaie, and F.~Sebastiano, ``A scalable cryo-cmos controller for the wideband frequency-multiplexed control of spin qubits and transmons,'' \emph{IEEE Journal of Solid-State Circuits}, vol.~55, pp. 2930--2946, 2020.

\bibitem{kang2022cryo}
K.~Kang, D.~Minn, S.~Bae, J.~Lee, S.~Bae, G.~Jung, S.~Kang, M.~Lee, H.-J. Song, and J.-Y. Sim, ``A cryo-cmos controller ic with fully integrated frequency generators for superconducting qubits,'' in \emph{2022 IEEE International Solid-State Circuits Conference (ISSCC)}, vol.~65.\hskip 1em plus 0.5em minus 0.4em\relax IEEE, 2022, pp. 362--364.

\bibitem{VanDijk2019a}
\BIBentryALTinterwordspacing
J.~P.~V. Dijk, E.~Kawakami, R.~N. Schouten, M.~Veldhorst, L.~M. Vandersypen, M.~Babaie, E.~Charbon, and F.~Sebastiano, ``Impact of classical control electronics on qubit fidelity,'' \emph{Physical Review Applied}, vol.~12, p.~1, 2019. [Online]. Available: \url{https://doi.org/10.1103/PhysRevApplied.12.044054}
\BIBentrySTDinterwordspacing

\bibitem{anders2023cmos}
J.~Anders, M.~Babaie, J.~C. Bardin, I.~Bashir, G.~Billiot, E.~Blokhina, S.~Bonen, E.~Charbon, J.~Chiaverini, I.~L. Chuang, C.~Degenhardt, D.~Englund, L.~Geck, L.~Le~Guevel, D.~Ham, R.~Han, M.~I. Ibrahim, D.~Krüger, K.~M. Lei, A.~Morel, D.~Nielinger, G.~Pillonnet, J.~M. Sage, F.~Sebastiano, R.~B. Staszewski, J.~Stuart, A.~Vladimirescu, P.~Vliex, and S.~P. Voinigescu, ``Cmos integrated circuits for the quantum information sciences,'' \emph{IEEE Transactions on Quantum Engineering}, vol.~4, pp. 1--30, 2023.

\bibitem{chow2021ibm}
J.~Chow, O.~Dial, and J.~Gambetta, ``Ibm quantum breaks the 100-qubit processor barrier,'' \emph{IBM Research Blog}, vol.~2, 2021.

\bibitem{storz2023loophole}
S.~Storz, J.~Sch{\"a}r, A.~Kulikov, P.~Magnard, P.~Kurpiers, J.~L{\"u}tolf, T.~Walter, A.~Copetudo, K.~Reuer, A.~Akin \emph{et~al.}, ``Loophole-free bell inequality violation with superconducting circuits,'' \emph{Nature}, vol. 617, no. 7960, pp. 265--270, 2023.

\bibitem{Ruf2021}
M.~Ruf, N.~H. Wan, H.~Choi, D.~Englund, and R.~Hanson, ``Quantum networks based on color centers in diamond,'' \emph{Journal of Applied Physics}, vol. 130, 2021.

\bibitem{pezzagna2021quantum}
S.~Pezzagna and J.~Meijer, ``Quantum computer based on color centers in diamond,'' \emph{Applied Physics Reviews}, vol.~8, no.~1, 2021.

\bibitem{Hensen2015}
B.~Hensen, H.~Bernien, A.~E. Dreaú, A.~Reiserer, N.~Kalb, M.~S. Blok, J.~Ruitenberg, R.~F. Vermeulen, R.~N. Schouten, C.~Abellán, W.~Amaya, V.~Pruneri, M.~W. Mitchell, M.~Markham, D.~J. Twitchen, D.~Elkouss, S.~Wehner, T.~H. Taminiau, and R.~Hanson, ``Loophole-free bell inequality violation using electron spins separated by 1.3 kilometres,'' \emph{Nature}, vol. 526, pp. 682--686, 10 2015.

\bibitem{Pompili2021}
\BIBentryALTinterwordspacing
M.~Pompili, S.~L.~N. Hermans, S.~Baier, H.~K.~C. Beukers, P.~C. Humphreys, R.~N. Schouten, R.~F.~L. Vermeulen, M.~J. Tiggelman, L.~Dos, S.~Martins, B.~Dirkse, S.~Wehner, and R.~Hanson, ``Realization of a multinode quantum network of remote solid-state qubits,'' pp. 259--264, 2021. [Online]. Available: \url{https://www.science.org}
\BIBentrySTDinterwordspacing

\bibitem{hermans2022qubit}
S.~Hermans, M.~Pompili, H.~Beukers, S.~Baier, J.~Borregaard, and R.~Hanson, ``Qubit teleportation between non-neighbouring nodes in a quantum network,'' \emph{Nature}, vol. 605, no. 7911, pp. 663--668, 2022.

\bibitem{Ishihara2021}
R.~Ishihara, J.~Hermias, S.~Yu, K.~Y. Yu, Y.~Li, S.~Nur, T.~Iwai, T.~Miyatake, K.~Kawaguchi, Y.~Doi, and S.~Sato, ``3d integration technology for quantum computer based on diamond spin qubits,'' \emph{Technical Digest - International Electron Devices Meeting, IEDM}, vol. 2021-Decem, pp. 14.5.1--14.5.4, 2021.

\bibitem{rugar2020generation}
A.~E. Rugar, H.~Lu, C.~Dory, S.~Sun, P.~J. McQuade, Z.-X. Shen, N.~A. Melosh, and J.~Vuckovic, ``Generation of tin-vacancy centers in diamond via shallow ion implantation and subsequent diamond overgrowth,'' \emph{Nano letters}, vol.~20, no.~3, pp. 1614--1619, 2020.

\bibitem{luo2022creation}
T.~Luo, L.~Lindner, J.~Langer, V.~Cimalla, X.~Vidal, F.~Hahl, C.~Schreyvogel, S.~Onoda, S.~Ishii, T.~Ohshima \emph{et~al.}, ``Creation of nitrogen-vacancy centers in chemical vapor deposition diamond for sensing applications,'' \emph{New Journal of Physics}, vol.~24, no.~3, p. 033030, 2022.

\bibitem{schirhagl2014nitrogen}
R.~Schirhagl, K.~Chang, M.~Loretz, and C.~L. Degen, ``Nitrogen-vacancy centers in diamond: nanoscale sensors for physics and biology,'' \emph{Annual review of physical chemistry}, vol.~65, pp. 83--105, 2014.

\bibitem{manson2006nitrogen}
N.~Manson, J.~Harrison, and M.~Sellars, ``Nitrogen-vacancy center in diamond: Model of the electronic structure and associated dynamics,'' \emph{Physical Review B}, vol.~74, no.~10, p. 104303, 2006.

\bibitem{iwasaki2017tin}
T.~Iwasaki, Y.~Miyamoto, T.~Taniguchi, P.~Siyushev, M.~H. Metsch, F.~Jelezko, and M.~Hatano, ``Tin-vacancy quantum emitters in diamond,'' \emph{Physical review letters}, vol. 119, no.~25, p. 253601, 2017.

\bibitem{rugar2019characterization}
A.~E. Rugar, C.~Dory, S.~Sun, and J.~Vu{\v{c}}kovi{\'c}, ``Characterization of optical and spin properties of single tin-vacancy centers in diamond nanopillars,'' \emph{Physical Review B}, vol.~99, no.~20, p. 205417, 2019.

\bibitem{thiering2018ab}
G.~Thiering and A.~Gali, ``Ab initio magneto-optical spectrum of group-iv vacancy color centers in diamond,'' \emph{Physical Review X}, vol.~8, no.~2, p. 021063, 2018.

\bibitem{iwasaki2020color}
T.~Iwasaki, ``Color centers based on heavy group-iv elements,'' in \emph{Semiconductors and Semimetals}.\hskip 1em plus 0.5em minus 0.4em\relax Elsevier, 2020, vol. 103, pp. 237--256.

\bibitem{doi2014deterministic}
Y.~Doi, T.~Makino, H.~Kato, D.~Takeuchi, M.~Ogura, H.~Okushi, H.~Morishita, T.~Tashima, S.~Miwa, S.~Yamasaki \emph{et~al.}, ``Deterministic electrical charge-state initialization of single nitrogen-vacancy center in diamond,'' \emph{Physical Review X}, vol.~4, no.~1, p. 011057, 2014.

\bibitem{gorlitz2022coherence}
J.~G{\"o}rlitz, D.~Herrmann, P.~Fuchs, T.~Iwasaki, T.~Taniguchi, D.~Rogalla, D.~Hardeman, P.-O. Colard, M.~Markham, M.~Hatano \emph{et~al.}, ``Coherence of a charge stabilised tin-vacancy spin in diamond,'' \emph{npj Quantum Information}, vol.~8, no.~1, p.~45, 2022.

\bibitem{Doherty2013}
\BIBentryALTinterwordspacing
M.~W. Doherty, N.~B. Manson, P.~Delaney, F.~Jelezko, J.~Wrachtrup, and L.~C. Hollenberg, ``The nitrogen-vacancy colour centre in diamond,'' \emph{Physics Reports}, vol. 528, pp. 1--45, 2013. [Online]. Available: \url{http://dx.doi.org/10.1016/j.physrep.2013.02.001}
\BIBentrySTDinterwordspacing

\bibitem{Debroux2021}
R.~Debroux, C.~P. Michaels, C.~M. Purser, N.~Wan, M.~E. Trusheim, J.~A. Martínez, R.~A. Parker, A.~M. Stramma, K.~C. Chen, L.~D. Santis, E.~M. Alexeev, A.~C. Ferrari, D.~Englund, D.~A. Gangloff, and M.~Atatüre, ``Quantum control of the tin-vacancy spin qubit in diamond,'' \emph{Physical Review X}, vol.~11, 12 2021.

\bibitem{acosta2013nitrogen}
V.~Acosta and P.~Hemmer, ``Nitrogen-vacancy centers: Physics and applications,'' \emph{MRS bulletin}, vol.~38, no.~2, pp. 127--130, 2013.

\bibitem{Bradley2019}
C.~E. Bradley, J.~Randall, M.~H. Abobeih, R.~C. Berrevoets, M.~J. Degen, M.~A. Bakker, M.~Markham, D.~J. Twitchen, and T.~H. Taminiau, ``A ten-qubit solid-state spin register with quantum memory up to one minute,'' \emph{Physical Review X}, vol.~9, 9 2019.

\bibitem{nguyen2019quantum}
C.~Nguyen, D.~Sukachev, M.~Bhaskar, B.~Machielse, D.~Levonian, E.~Knall, P.~Stroganov, R.~Riedinger, H.~Park, M.~Lon{\v{c}}ar \emph{et~al.}, ``Quantum network nodes based on diamond qubits with an efficient nanophotonic interface,'' \emph{Physical review letters}, vol. 123, no.~18, p. 183602, 2019.

\bibitem{mindarava2020synthesis}
Y.~Mindarava, R.~Blinder, Y.~Liu, J.~Scheuer, J.~Lang, V.~Agafonov, V.~A. Davydov, C.~Laube, W.~Knolle, B.~Abel \emph{et~al.}, ``Synthesis and coherent properties of 13c-enriched sub-micron diamond particles with nitrogen vacancy color centers,'' \emph{Carbon}, vol. 165, pp. 395--403, 2020.

\bibitem{woods1990nitrogen}
G.~Woods, G.~Purser, A.~Mtimkulu, and A.~Collins, ``The nitrogen content of type ia natural diamonds,'' \emph{Journal of Physics and Chemistry of Solids}, vol.~51, no.~10, pp. 1191--1197, 1990.

\bibitem{Yang2016}
\BIBentryALTinterwordspacing
W.~Yang, W.-L. Ma, and R.-B. Liu, ``Quantum many-body theory for electron spin decoherence in nanoscale nuclear spin baths,'' 7 2016. [Online]. Available: \url{http://arxiv.org/abs/1607.03993 http://dx.doi.org/10.1088/0034-4885/80/1/016001}
\BIBentrySTDinterwordspacing

\bibitem{kobayashi2020electrical}
S.~Kobayashi, Y.~Matsuzaki, H.~Morishita, S.~Miwa, Y.~Suzuki, M.~Fujiwara, and N.~Mizuochi, ``Electrical control for extending the ramsey spin coherence time of ion-implanted nitrogen-vacancy centers in diamond,'' \emph{Physical Review Applied}, vol.~14, no.~4, p. 044033, 2020.

\bibitem{wang2012comparison}
Z.-H. Wang, G.~De~Lange, D.~Rist{\`e}, R.~Hanson, and V.~Dobrovitski, ``Comparison of dynamical decoupling protocols for a nitrogen-vacancy center in diamond,'' \emph{Physical Review B}, vol.~85, no.~15, p. 155204, 2012.

\bibitem{childress2006coherent}
L.~Childress, M.~Gurudev~Dutt, J.~Taylor, A.~Zibrov, F.~Jelezko, J.~Wrachtrup, P.~Hemmer, and M.~Lukin, ``Coherent dynamics of coupled electron and nuclear spin qubits in diamond,'' \emph{Science}, vol. 314, no. 5797, pp. 281--285, 2006.

\bibitem{Trusheim2020}
M.~E. Trusheim, B.~Pingault, N.~H. Wan, M.~Gündoǧan, L.~D. Santis, R.~Debroux, D.~Gangloff, C.~Purser, K.~C. Chen, M.~Walsh, J.~J. Rose, J.~N. Becker, B.~Lienhard, E.~Bersin, I.~Paradeisanos, G.~Wang, D.~Lyzwa, A.~R. Montblanch, G.~Malladi, H.~Bakhru, A.~C. Ferrari, I.~A. Walmsley, M.~Atatüre, and D.~Englund, ``Transform-limited photons from a coherent tin-vacancy spin in diamond,'' \emph{Physical Review Letters}, vol. 124, 1 2020.

\bibitem{vanderSypen2005}
L.~M.~K. Vandersypen and I.~L. Chuang, ``Nmr techniques for quantum control and computation.''

\bibitem{10_qubit_array_sup}
``Supplemental material: A 10-qubit solid-state spin register with quantum memory up to one minute.''

\bibitem{Abobeih2018}
M.~H. Abobeih, J.~Cramer, M.~A. Bakker, N.~Kalb, M.~Markham, D.~J. Twitchen, and T.~H. Taminiau, ``One-second coherence for a single electron spin coupled to a multi-qubit nuclear-spin environment,'' \emph{Nature Communications}, vol.~9, 12 2018.

\bibitem{rosenthal2023microwave}
E.~I. Rosenthal, C.~P. Anderson, H.~C. Kleidermacher, A.~J. Stein, H.~Lee, J.~Grzesik, G.~Scuri, A.~E. Rugar, D.~Riedel, S.~Aghaeimeibodi \emph{et~al.}, ``Microwave spin control of a tin-vacancy qubit in diamond,'' \emph{Physical Review X}, vol.~13, no.~3, p. 031022, 2023.

\bibitem{thesis_connor}
C.~Bradley, ``Order from disorder: Control of multi-qubit spin registers in diamond,'' PhD thesis, Delft University of Technology, Delft, NL, October 2021, available at \url{https://doi.org/10.4233/uuid:acafe18b-3345-4692-9c9b-05e970ffbe40}.

\bibitem{thesis_pfaff}
W.~Pfaff, ``Quantum measurement and entanglement of spin quantum bits in diamond,'' PhD thesis, Delft University of Technology, Delft, NL, December 2013, available at \url{https://doi.org/10.4233/uuid:62208a4a-72c6-45b5-af51-1a8e8f8bfdd8}.

\bibitem{taminiau2014universal}
T.~H. Taminiau, J.~Cramer, T.~van~der Sar, V.~V. Dobrovitski, and R.~Hanson, ``Universal control and error correction in multi-qubit spin registers in diamond,'' \emph{Nature nanotechnology}, vol.~9, no.~3, pp. 171--176, 2014.

\bibitem{hopper2018spin}
D.~A. Hopper, H.~J. Shulevitz, and L.~C. Bassett, ``Spin readout techniques of the nitrogen-vacancy center in diamond,'' \emph{Micromachines}, vol.~9, no.~9, p. 437, 2018.

\bibitem{schwartz2018robust}
I.~Schwartz, J.~Scheuer, B.~Tratzmiller, S.~M{\"u}ller, Q.~Chen, I.~Dhand, Z.-Y. Wang, C.~M{\"u}ller, B.~Naydenov, F.~Jelezko \emph{et~al.}, ``Robust optical polarization of nuclear spin baths using hamiltonian engineering of nitrogen-vacancy center quantum dynamics,'' \emph{Science advances}, vol.~4, no.~8, p. eaat8978, 2018.

\bibitem{bernien2013heralded}
H.~Bernien, B.~Hensen, W.~Pfaff, G.~Koolstra, M.~S. Blok, L.~Robledo, T.~H. Taminiau, M.~Markham, D.~J. Twitchen, L.~Childress \emph{et~al.}, ``Heralded entanglement between solid-state qubits separated by three metres,'' \emph{Nature}, vol. 497, no. 7447, pp. 86--90, 2013.

\bibitem{sipahigil2014indistinguishable}
A.~Sipahigil, K.~D. Jahnke, L.~J. Rogers, T.~Teraji, J.~Isoya, A.~S. Zibrov, F.~Jelezko, and M.~D. Lukin, ``Indistinguishable photons from separated silicon-vacancy centers in diamond,'' \emph{Physical review letters}, vol. 113, no.~11, p. 113602, 2014.

\bibitem{VanDijk2020}
J.~P.~V. Dijk, B.~Patra, S.~Pellerano, E.~Charbon, F.~Sebastiano, and M.~Babaie, ``Designing a dds-based soc for high-fidelity multi-qubit control,'' \emph{IEEE Transactions on Circuits and Systems I: Regular Papers}, vol.~67, pp. 5380--5393, 12 2020.

\bibitem{metsch2019initialization}
M.~H. Metsch, K.~Senkalla, B.~Tratzmiller, J.~Scheuer, M.~Kern, J.~Achard, A.~Tallaire, M.~B. Plenio, P.~Siyushev, and F.~Jelezko, ``Initialization and readout of nuclear spins via a negatively charged silicon-vacancy center in diamond,'' \emph{Physical review letters}, vol. 122, no.~19, p. 190503, 2019.

\bibitem{kalb2018dephasing}
N.~Kalb, P.~C. Humphreys, J.~Slim, and R.~Hanson, ``Dephasing mechanisms of diamond-based nuclear-spin memories for quantum networks,'' \emph{Physical Review A}, vol.~97, no.~6, p. 062330, 2018.

\bibitem{de2010universal}
G.~De~Lange, Z.-H. Wang, D.~Riste, V.~Dobrovitski, and R.~Hanson, ``Universal dynamical decoupling of a single solid-state spin from a spin bath,'' \emph{Science}, vol. 330, no. 6000, pp. 60--63, 2010.

\bibitem{Green2013}
T.~J. Green, J.~Sastrawan, H.~Uys, and M.~J. Biercuk, ``Arbitrary quantum control of qubits in the presence of universal noise,'' \emph{New Journal of Physics}, vol.~15, 9 2013.

\bibitem{chang2021detecting}
J.~Chang, J.~Los, J.~Tenorio-Pearl, N.~Noordzij, R.~Gourgues, A.~Guardiani, J.~Zichi, S.~Pereira, H.~Urbach, V.~Zwiller \emph{et~al.}, ``Detecting telecom single photons with 99.5- 2.07+ 0.5\% system detection efficiency and high time resolution,'' \emph{APL Photonics}, vol.~6, no.~3, 2021.

\bibitem{chang2019multimode}
J.~Chang, I.~E. Zadeh, J.~W. Los, J.~Zichi, A.~Fognini, M.~Gevers, S.~Dorenbos, S.~F. Pereira, P.~Urbach, and V.~Zwiller, ``Multimode-fiber-coupled superconducting nanowire single-photon detectors with high detection efficiency and time resolution,'' \emph{Applied optics}, vol.~58, no.~36, pp. 9803--9807, 2019.

\bibitem{ravindran2020active}
P.~Ravindran, R.~Cheng, H.~Tang, and J.~C. Bardin, ``Active quenching of superconducting nanowire single photon detectors,'' \emph{Optics express}, vol.~28, no.~3, pp. 4099--4114, 2020.

\bibitem{bogaerts2020programmable}
W.~Bogaerts, D.~P{\'e}rez, J.~Capmany, D.~A. Miller, J.~Poon, D.~Englund, F.~Morichetti, and A.~Melloni, ``Programmable photonic circuits,'' \emph{Nature}, vol. 586, no. 7828, pp. 207--216, 2020.

\bibitem{seok2016large}
T.~J. Seok, N.~Quack, S.~Han, R.~S. Muller, and M.~C. Wu, ``Large-scale broadband digital silicon photonic switches with vertical adiabatic couplers,'' \emph{Optica}, vol.~3, no.~1, pp. 64--70, 2016.

\bibitem{Sukachev2017}
D.~D. Sukachev, A.~Sipahigil, C.~T. Nguyen, M.~K. Bhaskar, R.~E. Evans, F.~Jelezko, and M.~D. Lukin, ``Silicon-vacancy spin qubit in diamond: A quantum memory exceeding 10 ms with single-shot state readout,'' \emph{Physical Review Letters}, vol. 119, pp. 1--6, 2017.

\bibitem{pryor2009multiphysics}
R.~W. Pryor, \emph{Multiphysics modeling using COMSOL{\textregistered}: a first principles approach}.\hskip 1em plus 0.5em minus 0.4em\relax Jones \& Bartlett Publishers, 2009.

\bibitem{Lacy2020}
J.~H. Lacy, A.~Cridland, J.~Pinder, A.~Uribe, R.~Willetts, and J.~Verdu, ``Superconducting flux pump for a planar magnetic field source,'' \emph{IEEE Transactions on Applied Superconductivity}, vol.~30, 12 2020.

\bibitem{patra2017cryo}
B.~Patra, R.~M. Incandela, J.~P. Van~Dijk, H.~A. Homulle, L.~Song, M.~Shahmohammadi, R.~B. Staszewski, A.~Vladimirescu, M.~Babaie, F.~Sebastiano \emph{et~al.}, ``Cryo-cmos circuits and systems for quantum computing applications,'' \emph{IEEE Journal of Solid-State Circuits}, vol.~53, no.~1, pp. 309--321, 2017.

\bibitem{hossen2019power}
M.~O. Hossen, B.~Chava, G.~Van~der Plas, E.~Beyne, and M.~S. Bakir, ``Power delivery network (pdn) modeling for backside-pdn configurations with buried power rails and $\backslash \mu$ tsvs,'' \emph{IEEE Transactions on Electron Devices}, vol.~67, no.~1, pp. 11--17, 2019.

\bibitem{dataset}
L.~Enthoven. (2024) \BIBforeignlanguage{English}{Data underlying the publication: Optimizing the electrical interface for large-scale color-center quantum processors}. [Online], Available: \url{http://dx.doi.org/10.4121/36f89257-d7dd-4b42-b032-2b0139acd538}.

\end{thebibliography}

\end{document}